\documentclass[apj,twocolumn, twocolappendix]{openjournal}
\usepackage{orcidlink}  

\usepackage{multirow}
\usepackage[parfill]{parskip}
\usepackage{graphicx}
\usepackage{amssymb}
\usepackage{amsmath}
\usepackage{epstopdf}
\usepackage{xcolor}

\usepackage{hyperref}
\hypersetup{
     colorlinks   = true,
     citecolor    = blue,
     linkcolor = blue,
     urlcolor = blue
}

\def\apj{{ ApJ}}
\def\apjl{{ApJL}}
\def\apjs{{ ApJS}}

\def\aap{{ A\&A}}

\def\mnras{{ MNRAS}}
\def\araa{{ ARA\&A}}

\def\nat {{ Nature}}
\def\pasj{{ PASJ}}

\def\prl{{ Phys. Rev. Lett}}

\def\nar{{New Astronomy Reviews}}

\def\mr{\mathrm}
\def\d{\mr{d}}

\def\mc{\mathcal}

\def\Msun{M_{\odot}}

\def\eps{\epsilon}

\def\me{m_{\rm e}}
\def\mproton{m_{\rm p}}
\def\cs{c_{\rm s}}
\def\kB{k_{\rm B}}

\def\rp{r_{\rm p}}
\def\rB{r_{\rm B}}
\def\rA{r_{\rm A}}
\def\rlc{r_{\rm lc}}
\def\rco{r_{\rm co}}
\def\rd{r_{\rm d}}
\def\rg{r_{\rm g}}
\def\Mns{M_{\rm ns}}
\def\muB{\mu_{\rm B}}
\def\cs{c_{\rm s}}

\newcommand{\lrb}[1]{\left({#1}\right)}
\newcommand{\lrsb}[1]{\left[{#1}\right]}

\newcommand{\abslt}[1]{\left|#1\right|}

\def\msunyr{M_\odot\mr{\,yr^{-1}}}

\def\checkmark{\tikz\fill[scale=0.4](0,.35) -- (.25,0) -- (1,.7) -- (.25,.15) -- cycle;}

\begin{document}


\title{Accretion from a shock-inflated companion:\\ double-peaked supernova lightcurve with periodic modulations}

\author{\vspace{-1.5cm}Wenbin Lu\,\orcidlink{0000-0002-1568-7461}$^{1}$, Savannah Cary\, \orcidlink{0000-0003-1860-1632}$^{1}$
and Daichi Tsuna\,\orcidlink{0000-0002-6347-3089}$^{2,3}$}

\affiliation{$^1$Department of Astronomy and Theoretical Astrophysics Center, University of California, Berkeley, CA 94720-3411, USA}
\affiliation{$^2$TAPIR, Mailcode 350-17, California Institute of Technology, Pasadena, CA 91125, USA}
\affiliation{$^3$Research Center for the Early Universe (RESCEU), School of Science, The University of Tokyo,  Bunkyo-ku, Tokyo 113-0033, Japan}
\email{wenbinlu@berkeley.edu, scary@berkeley.edu, tsuna@caltech.edu}

\begin{abstract}

We study the observational signatures from the interactions between a newly born neutron star and a companion star that is impacted by the supernova ejecta. We focus on the cases with bound post-explosion orbits, where the neutron star may periodically gravitationally capture gas from the companion. We find that neutron star accretion must occur if the pre-supernova binary separation is less than about $20R_\odot$. This is because the stellar radius expands beyond this radius before the shock-inflated envelope undergoes Kelvin-Helmholtz contraction back to the main sequence. We then consider the internal shocks formed between adjacent episodes of disk wind. The shocks efficiently convert the wind kinetic energy into radiation (due to inverse-Compton cooling), which heats up the supernova ejecta located at much larger radii. The extra heating powers bright optical emission that is periodically modulated on the orbital timescale. The shocks also accelerate non-thermal particles which produce $\gamma$-ray and neutrino emission from 100 MeV to 10 PeV via hardronic $pp$ collisions. The high-energy photons leak out of the supernova ejecta after a delay of several months to one year. Photo-ionization of the slowest parts of the disk wind produces hydrogen recombination lines. We then use the model to explain the puzzling Type Ic supernova SN2022jli which shows a double-peaked optical lightcurve along with many peculiar properties, including delayed onset and rapid shutoff of the second peak, periodic modulations, delayed GeV emission, and narrow Balmer lines.
Under this model, SN2022jli had a close companion at a pre-supernova binary separation of $10$--$20 R_\odot$, likely due to an earlier phase of common-envelope evolution.


\end{abstract}

\maketitle  

\section{Introduction}\label{sec:intro}

Progenitors of hydrogen-poor type Ib/c supernovae \citep{Filippenko:1997aa} are believed to be dominated by intermediate-mass ($\sim$2.5 to 8$\,M_\odot$) helium stars whose hydrogen envelopes had been stripped by binary interaction \citep[e.g.,][]{Podsiadlowski:1992aa, Woosley:1995aa, Eldridge:2008aa, Eldridge:2013aa, Smith:2011aa, Drout:2011aa, Taddia:2015aa, Taddia:2018ab, Yoon:2015aa, Lyman:2016aa, Prentice:2019aa, Dessart:2020aa, Ertl:2020aa, Woosley:2021aa}. In this picture, the fact that roughly 20\% of all core-collapse supernovae are type Ib/c explosions \citep[e.g.,][]{Li:2011aa, Eldridge:2013aa, Shivvers:2017aa} suggests that their companions are mostly main-sequence stars \citep[e.g.,][]{Liu:2015aa, Zapartas:2017aa}. 

In such binary systems, an important effect is the impact of the supernova ejecta on the companion star. This hydrodynamic interaction has a few interesting outcomes that are well-studied in the literature \citep[e.g.,][]{Wheeler:1975aa, Fryxell:1981aa, Marietta:2000aa, Kasen:2010aa, Hirai:2014aa, Rimoldi:2016aa, Hirai:2018aa, Liu:2015aa, Ogata:2021aa, Hober:2022aa, Chen:2023aa}: (i) A small fraction of the companion's mass is stripped;
(ii) the shock-heated bound layers of the companion star expands to much larger radii and the star may stay highly inflated for up to many decades after the impact; (iii) the momentum impact on the companion star may affect the orbital dynamics; (iv) the reverse shock that heats up part of the ejecta may briefly brighten up the supernova emission. These outcomes strongly depend on the mass of the companion star and the orbital separation, which are theoretically uncertain and observationally poorly constrained.

For a few type Ib/c supernovae, the post-explosion images provide potential detections or stringent upper limits on the luminosity of the companion stars \citep{Van-Dyk:2016aa, Folatelli:2016aa, Eldridge:2016aa, Maund:2016aa, Sun:2020aa, Fox:2022aa}. Specifically, in the case of SN2006jc, observations at 3.6 and 10.4 years after the supernova unveil a point source whose spectral energy distribution is roughly consistent with a shock-inflated companion \citep{Maund:2016aa, Sun:2020aa, Ogata:2021aa, Chen:2023aa}. 

Another independent approach is to search for and study pre-explosion helium star-main sequence (He-MS) binaries in the Milky Way or nearby galaxies.  \citet{Drout:2023aa} identified a sample of 25 objects based on UV photometric excess of stars in the Large/Small Magellanic Clouds \citep[more recently,][selected a larger sample]{ludwig25_stripped_stars}. Eight of the 25 objects appear as single-lined spectroscopic binaries with the optical spectrum dominated by the He star (their ``Class 1''). Spectroscopic modeling by \cite{Drout:2023aa} shows that the system only appears as ``Class 1'' if the \textit{current} mass ratio between the MS star and He star is sufficiently small ($q \lesssim 0.6$). Since the He star's mass is only a small fraction (roughly 1/3 for the relevant masses) of the initial mass at zero-age main sequence, \cite{Gotberg:2023aa} suggest that the earlier stage of Case B mass transfer that led to the stripping of the He star was dynamically unstable because of the very low accretor-to-donor mass ratio ($q\lesssim 0.2$) back then. The authors argue that a fraction of He-MS binaries are the products of common-envelope evolution, so we expect the two stars in these systems to be closely separated (the binary separation can be constrained by the not-yet-published radial velocity measurements).

Based on the mass-luminosity relation of core He burning stars, \cite{Gotberg:2023aa} roughly estimate the ``evolutionary masses'' of the He stars in the ``Class 1'' sample to be in the range $2\lesssim M_{\rm He} \lesssim 8M_\odot$ with a median around $3.3M_\odot$.
Thus, it is likely that more than half of these He stars will later on produce core-collapse supernovae \citep{Woosley:2019aa}. Such supernovae would belong to Type Ib/c, if the remaining hydrogen-rich layer and a fraction of the helium-rich layer are further stripped by wind and mass transfer \citep[e.g.,][]{Tauris:2015aa, Laplace:2020aa}. If there is a substantial amount of hydrogen left at explosion, then the supernova would belong to Type IIb, which is another type of stripped-envelope supernovae that we do not discuss in this paper. In those cases, the companion star may be in orbits that are too wide for the ejecta-companion interaction to be important.

Motivated by the above observations, we \textit{assume} that a fraction of the progenitors of Type Ib/c supernovae are in close orbits with a MS companion star. For such close orbits, the newly born neutron star may remain bound to the companion star even for moderately strong natal kicks. Therefore, if the shock-heated outer layers of the companion star expands to large radii, the bound neutron star will likely strongly interact with the inflated stellar envelope. Such interactions have been predicted by \citet{Rimoldi:2016aa, Hirai:2018aa, Ogata:2021aa, Hober:2022aa}. These authors briefly considered the possibility of mass transfer, neutron star accretion, orbital circularization, and formation of Thorne-\.Zytkow objects \citep{Thorne:1977aa} and pulsar planets. \citet{Hirai:2022aa} carried out hydrodynamic simulations of very close encounters between a neutron star and an \textit{unperturbed} companion star and discussed more effects such as high-velocity ejection of the companion star,
merger and tidal capture, and super-Eddington accretion onto the neutron star.

In this paper, we study in more detail the interactions between the neutron star and the shock-inflated companion star, with the goal of predicting the observational signatures of such interactions during or shortly after the supernova explosion. This is similar to the work of \citet{tsuna25_TDEs}, where rarer cases of the tidal disruption of the companion star are considered. We focus on the cases with bound\footnote{In the unbound cases, the neutron star can still capture up to $\sim\! 10^{-4}M_\odot$ \citep{cary25_ULP}, which can provide additional power to the supernova emission.} post-explosion orbits, where the neutron star can periodically gravitationally capture gas from the companion star's outer envelope --- as the stellar envelope expands significantly after the shock impact, these cases are much more common than tidal disruptions. Our consideration is motivated by the recent discovery of a Type Ic supernova SN2022jli \citep{Chen:2024aa, Moore:2023aa, cartier24_SN2022jli}, which shows a double-peaked lightcurve along with a number of peculiar properties during the second peak such as periodic modulations, GeV emission, and narrow Balmer emission lines. It is shown that the the second peak of the lightcurve is consistent with being powered by neutron star accretion from the temporarily inflated envelope of the companion star.

This paper is organized as follows. In \S \ref{sec:entropy_jump}, we develop a semi-analytic formalism to study the shock-heating of the outer layers of the companion star. In \S \ref{sec:MESA_runs}, we follow the stellar evolution after the shock impact. In \S \ref{sec:NS_accretion}, we study the gas accretion onto the neutron star and the associated energy output. In \S \ref{sec:shock_emission}, we consider the emission from the internal shocks formed between adjacent episodes of accretion disk wind. In \S \ref{sec:application_SN2022jli}, we apply our model to the case of SN2022jli. Finally, our results are summarized in \S \ref{sec:summary}.
Throughout this paper, we will take the neutron star's mass, radius, and moment of inertia to be $M_{\rm ns} = 1.4M_\odot$, $R_{\rm ns} = 12\rm\,km$, and $I_{\rm ns} = 1.3\times10^{45}\rm\, g\,cm^2$, respectively. We use CGS units.

\section{Entropy jump due to shock heating of the stellar envelope}\label{sec:entropy_jump}

Let us consider an unperturbed main-sequence companion star of mass $M_*$ and radius $R_*$.  The unperturbed stellar density and pressure profiles are denoted as $\rho(m)$ and $P(m)$, where $m$ is the enclosed mass coordinate. The pre-explosion orbit is assumed to be circular and the separation between the supernova progenitor and the companion star is denoted as $a_0$, which is generally much larger than the stellar radius.


At the position of the star, the unperturbed supernova ejecta has density $\rho_{\rm ej}(t)$ and velocity $v_{\rm ej}(t)$ at time $t$ since the explosion. The ejecta provides a ram pressure $P_{\rm ram}(t)=\rho_{\rm ej} v_{\rm ej}^2$ on one side of the star. For an explosion with ejecta energy $E_{\rm ej}$ and mass $M_{\rm ej}$, we can estimate the characteristic ejecta velocity as $v_{\rm ej} \simeq \sqrt{2E_{\rm ej}/M_{\rm ej}}$, and the ejecta's density when it reaches the star's position is roughly given by $\rho_{\rm ej} \simeq 3M_{\rm ej}/(4\pi a_0^3)$. Thus, the typical ram pressure of the ejecta can be estimated as
\begin{equation}\label{eq:ram_pres}
    P_{\rm ram} \simeq {3E_{\rm ej}\over 2\pi a_0^3} = 1.4\times10^{15}\mr{erg\over cm^{3}}\,
    E_{\rm ej,51} \lrb{ 10R_\odot \over a_0}^{3},
\end{equation}
where $E_{\rm ej,51} = E_{\rm ej}/10^{51}\rm\, erg$.
Comparing eq. (\ref{eq:ram_pres}) with the average pressure inside the star
\begin{equation}\label{eq:Pbar}
    \bar{P} = {GM_*^2\over 4\pi R_*^4} = 8.9\times10^{14}\mr{erg\over cm^{3}} \lrb{M_*\over M_\odot}^2 \lrb{R_\odot\over R_*}^{4},
\end{equation}
we see that, for a companion star in a very close orbit with $a_0\lesssim 10R_\odot$, the forward shock will heat up a significant fraction of the star's mass; for wide orbits with $a_0\gtrsim 10^2R_\odot$, the forward shock only heats up a very small fraction of the star's mass. In this paper, we focus on the less extreme cases with $a_0\gtrsim 10R_\odot$.

\begin{figure}
\centering
\includegraphics[width=0.47\textwidth]{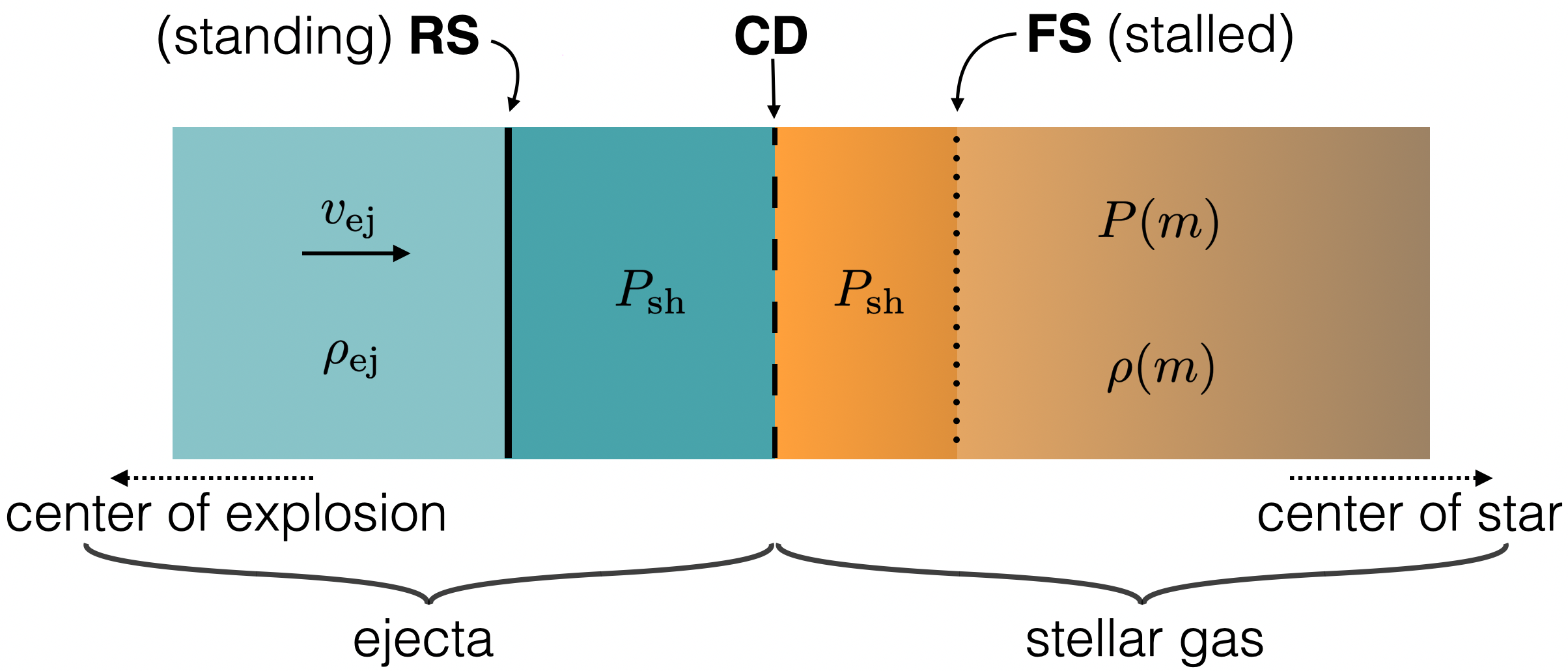}
\caption{Simplified 1D \textit{hydrostatic} configuration of the interaction between the supernova ejecta (blue regions) and the star (orange regions). In this picture, the ejecta provides a quasi-steady ram pressure $P_{\rm ram} = \rho_{\rm ej} v_{\rm ej}^2$, which initially drives a forward shock (FS) into the star. As the FS propagates to a layer of mass coordinate $m$ where the unperturbed pressure $P(m)$ is comparable to the ran pressure $P_{\rm ram}$, the FS stalls and does not heat up the deeper layers. There is also a reverse shock (RS) that heats up the ejecta. Between the FS and the RS, the shock-heated ejecta and stellar gas meets each other at the contact discontinuity (CD). The pressure of the shocked region between the FS and the RS is uniform and denoted as $P_{\rm sh}$.
}
\label{fig:shock_structure}
\end{figure}

The ram pressure varies on a timescale of
\begin{equation}
    t_{\rm ram} \sim {a_0\over v_{\rm ej}} = 1.2\times10^3\mr{\,s}\, {a_0\over 10R_\odot} E_{\rm ej,51}^{-1/2} \lrb{M_{\rm ej}\over 3M_\odot}^{1/2},
\end{equation}
where we have taken a fiducial ejecta mass for Type Ib/c supernovae. This should be compared with the shock-crossing time $t_{\rm sh}$ from the surface of the star to the depth where the forward shock stalls. Since the depth of the shock-heated layer is less than or equal to the stellar radius and the speed of the forward shock is greater than the local sound speed, we know that the shock-crossing time $t_{\rm sh}$ is generally shorter than the dynamical timescale $t_*$ of the unperturbed star,
\begin{equation}
    t_{\rm sh} < t_{*} = \sqrt{R_*^3\over GM_*} = 1.6\times10^3\mr{\,s}\, {(R_*/R_\odot)^{3/2}\over (M_*/M_\odot)^{1/2}}.
\end{equation}
The above equations show that $t_{\rm ram}$ is slightly larger than but comparable to $t_{\rm sh}$. This motivates us to consider the structure of the shock-heated outer layers of the star to be in quasi-hydrostatic equilibrium. Such an equilibrium is schematically shown in Fig. \ref{fig:shock_structure}, where we have simplified the structure of the shocked gas into a one-dimensional (1D) picture.



After the forward shock sweeps past a given layer at mass coordinate $m$, the pressure of the shock-heated region is raised to $P_{\rm sh}$, which is assumed to be roughly equal to the ram pressure $P_{\rm ram}$ provided by the ejecta,
\begin{equation}
    P_{\rm sh}\simeq P_{\rm ram}.
\end{equation}
The Mach number of the upstream gas in the frame of the forward shock as it passes mass coordinate $m$ is given by (from the Rankine-Hugoniot jump conditions)
\begin{equation}\label{eq:Mach_number}
    \mc{M}^2(m) = 1 + {\gamma + 1\over 2\gamma} \lrsb{{P_{\rm sh}\over P(m)} -1},
\end{equation}
where $\gamma=5/3$ is the adiabatic index. Hereafter, we assume that gas pressure dominates over radiation pressure in the shocked stellar gas, which is true in the deeper layers where most of the energy associated with the shock heating is deposited, even though radiation pressure may become important in the outermost shock-heated layers of the star (where only a negligible fraction of the heat is deposited).

The density and temperature of the shock-heated stellar gas are given by
\begin{equation}\label{eq:density_jump}
    {\rho_{\rm sh}(m)\over \rho(m)} = {(\gamma+1)\mc{M}^2\over (\gamma-1)\mc{M}^2 + 2} = {4P_{\rm sh}/P + 1 \over 4 + P_{\rm sh}/P},
\end{equation}
and
\begin{equation}\label{eq:temperature_of_shocked_gas}
    T_{\rm sh}(m) = {\mu \mproton P_{\rm sh}\over \kB \rho_{\rm sh}(m)},
\end{equation}
where $\kB$ is the Boltzmann constant and $\mu\approx 0.6$ is the mean molecular weight for solar composition.




For a non-relativistic Maxwell-Boltzmann gas that is a mixture of $i=1,2,\ldots$ species at the same temperature $T$, the entropy per particle $s_i$ for each species with number density $n_i$ and particle mass $m_i$ is given by
\begin{equation}\label{eq:entropy}
    {s_i\over \kB} = {5\over 2} + \ln\lrsb{{g_i\over n_i}\lrb{2\pi m_i \kB T\over h^2}^{3/2}},
\end{equation}
where $g_i$ is the statistical weight: $g_i=2$ for protons and electrons (spin 1/2 particles) whereas $g_i=1$ for helium nuclei (spin 0).
The total entropy per unit mass (or specific entropy) of the gas mixture is given by
\begin{equation}\label{eq:total_entropy}
    s = {\sum_i n_i s_i\over \rho}.
\end{equation}
As long as the gas composition stays unchanged across the shock front, the specific entropy has the following dependences on pressure and density:
\begin{equation}\label{eq:total_entropy_P_rho}
     s={3\kB \over 2 \mu \mproton}\ln(P/\rho^{5/3})+\mr{const}.
\end{equation}

Based on eqs. (\ref{eq:total_entropy_P_rho}) and (\ref{eq:density_jump}), we can calculate the entropy jump due to the heating by the forward shock $\Delta s(m) = s_{\rm sh} - s$, where $s(m)$ and $s_{\rm sh}(m)$ are the entropies for the unperturbed and shock-heated gas at mass coordinate $m$, respectively. The result is
\begin{equation}\label{eq:entropy_jump}
\begin{split}
 \Delta s(m) &= {\kB\over \mu \mproton} \lrsb{{3\over 2}\ln{P_{\rm sh}\over P(m)} - {5\over 2}\ln {\rho_{\rm sh}\over \rho(m)}} \\
    &={3\kB \over 2\mu \mproton}\ln\lrsb{x\lrb{4+x\over 1+4x}^{5/3}}, \ \ x \equiv {P_{\rm sh}\over P(m)}.
\end{split}
\end{equation}
If the ram pressure is much smaller than the central pressure of the star, then there is a minimum mass coordinate $m_{\rm min}$ at which $\Delta s(m_{\rm min})=0$ and $P_{\rm sh}=P(m_{\rm min})$. This mass coordinate is where the forward shock stalls. If the ram pressure is higher than the central pressure of the star, then the entire star will be shock heated in our quasi-static picture.

The pressure profile is the simplest for the outermost layers of the star where the enclosed mass is $m \approx M_*$ (and $r\approx R_*$). Based on the equation of hydrostatic equilibrium (assuming a non-rotating star),
\begin{equation}
    {\d P\over \d r} = - {Gm(r)\over r^2} \rho = -{Gm(r)\over r^2} {\d m\over 4\pi r^2 \d r},
\end{equation}
we take $r\approx R_*$ and $m\approx M_*$ and obtain
\begin{equation}
    \d P \approx {GM_*\over 4\pi R_*^4} \d m.
\end{equation}
This means that the pressure is related to the enclosed mass $m$ via
\begin{equation}\label{eq:outer_layers_pressure_profile}
    P(m) \approx \bar{P} \lrb{1-m/M_*},\ \mbox{ for } m\approx M_*,
\end{equation}
where $\bar{P}$ is given by eq. (\ref{eq:Pbar}). After the passage of the forward shock, the pressure jump at mass coordinate $m$ is given by
\begin{equation}
    x(m)={P_{\rm sh}\over P(m)} \approx {P_{\rm sh} \over \bar{P}} {M_*\over M_* - m}, 
\end{equation}
which can then be plugged into eq. (\ref{eq:entropy_jump}) to obtain the entropy jump. In the limit of $x\gg 1$ near the stellar surface, the entropy jump is given by
\begin{equation}\label{eq:asymptotic_entropy_jump}
    \Delta s(m) \approx {3\kB \over 2 \mu \mproton} \ln \lrsb{{P_{\rm sh}\over 4^{5/3} \bar{P}} {M_*\over M_* - m}}, \mbox{ for } m\approx M_*.
\end{equation}
We see that the entropy jump has the following scaling $\mr{e}^{\Delta s(m)}\propto (M_*-m)^{-1}$, which is in agreement with the results from numerical simulations by \citet[][see their Fig. 18]{Hirai:2018aa}. Our analytical theory also predicts that the entropy jump diverges logarithmically near the surface of the star where $m/M_*\rightarrow 1$. This divergence is unphysical because (i) the layers very close to the stellar surface will be rapidly stripped away from the star (hydrostatic equilibrium is not possible), and (ii) the surface layers at sufficiently small optical depth can radiatively cool. In the Appendix, we show that this divergence is unimportant for the subsequent evolution of the ejecta-impacted companion star, because most of the shock-generated heat is deposited near the deepest layer where the forward shock stalls.

In our 1D hydrostatic picture, we assume $P_{\rm sh} \simeq P_{\rm ram}$.
However, realistic systems are different from our simplified picture because (i) both the forward and reverse shocks are generally oblique mainly due to the curvature of the stellar surface and (ii) the shocked gas can escape sideways which will reduce $P_{\rm sh}$ as compared to the 1D hydrostatic case. Therefore, the exact ratio of $P_{\rm sh}/P_{\rm ram}$ needs to be calibrated against numerical simulations in future works for the purpose of making more accurate observational predictions.

Apart from the 1D simplification, another shortcoming of our analytic model is the ignorance of mass loss from the companion star. It is well-known that the ejecta's impact causes the highest entropy surface layers of the star to become unbound to the star \citep{Wheeler:1975aa}. This would lead to the flattening of the entropy profile of the bound layers at very small exterior masses. This flattening is not captured in our analytic model. However, the flattening of the entropy profile makes very little practical difference in the late-time evolution of the star for the following reason. Most of the shock heating associated with the entropy jump is deposited near the bottom of the shocked stellar envelope where $P(m) \sim P_{\rm sh}$. Only a very small fraction of the heat is deposited to the surface layers that escape from the star and this amount of heat is quickly radiated away in the subsequent evolution of the shock-heated star. The unimportance of the entropy profile flattening is demonstrated in the Appendix (see Fig. \ref{fig:test_ds_flattening}). Since the fractional mass loss is a steeply decreasing function of the separation between the two stars \citep[e.g.,][]{Hirai:2018aa}, we expect our analytic model to work best for relatively large binary separations.

In the next section, we take the entropy jump given by our analytic model (eq. \ref{eq:entropy_jump}) as the initial conditions to study the subsequent evolution of the shock-heated stars. 


\section{Evolution of shock-heated stars}\label{sec:MESA_runs}


In this section, we use Modules for Experiments in Stellar Astrophysics
\citep[MESA,][]{Paxton2011, Paxton2013, Paxton2015, Paxton2018, Paxton2019, Jermyn2023} to study the long-term evolution of the shock-heated companion star. We apply the model to a large grid of parameters $(M_*, a_0)$ that would be more computationally expensive for detailed multi-dimensional numerical simulations to cover. Our method, which follows the work by \citet{Bauer:2019aa} \citep[see also][]{Pan:2012aa}, is described in \S \ref{sec:MESA_method}, and our results are presented in \S \ref{sec:MESA_results}.

\subsection{Method}\label{sec:MESA_method}

The initial condition is a zero-age main-sequence star of a given mass $M_*$ at solar metallicity. 
To capture the effect of the ejecta's impact, we model the entropy jump as given by eq. (\ref{eq:entropy_jump}) at each mass coordinate $m$ by heat injection at a rate $\dot{q}(m, t)$ (heating rate per unit mass) such that
\begin{equation}\label{eq:heating_rate}
    \Delta s(m) = \int_0^{\Delta t}  {\dot{q}(m, t) \over T(m, t)} \d t,
\end{equation}
where $\Delta t$ is the heating duration and $T(m,t)$ is the current temperature profile at time $t$ since the beginning of heating.
We pick a global duration $\Delta t$ and then adopt a steady entropy injection rate
\begin{equation}
    \dot{s}(m) = {\Delta s(m)/\Delta t},
\end{equation}
which is applied to each Lagrangian layer as labeled by its mass coordinate $m$. Note that the heating rate at a given time is adjusted according to the current temperature $T(m, t)$ such that it produces the desired constant entropy injection rate $\dot{s}(m)$. At $t > \Delta t$ since the beginning of heating, we turn off heating and continue to evolve the star for a maximum age of $10^4\rm\, yr$ by which time the star has returned to the main-sequence. In all our simulations, we fix $\Delta t=10^4\rm\, s$ (roughly 0.1 d), as our results are insensitive to $\Delta t$ as long as it is much shorter than the thermal expansion timescale of the envelope (this is tested in the Appendix). 





To avoid the logarithmic divergence of the entropy jump near the stellar surface, we consider a minimum \textit{exterior} mass $m_{\rm ex,min}$ below which the entropy jump is assumed to be flat, meaning that
\begin{equation}\label{eq:mex_min}
    \Delta s(m>M_*-m_{\rm ex,min}) = \Delta s(m=M_*-m_{\rm ex,min}).
\end{equation}
In all our simulations, we fix $m_{\rm ex,min}=10^{-5}M_\odot$ and the evolution of the shock-heated star is insensitive to our choice of $m_{\rm ex,min}$ as long as it is much smaller than the total mass of the shock-heated layers (this is tested in the Appendix).

In our model, the ejecta energy $E_{\rm ej}$ is degenerate with the pre-explosion orbital separation $a_0$ through the ram pressure in eq. (\ref{eq:ram_pres}), so we fix $E_{\rm ej} = 10^{51}\rm\, erg$ as is typical for Type Ib/c supernovae \citep[e.g.,][]{Lyman:2016aa}.



Then, we are left with two free parameters, the companion mass $M_*$ and pre-explosion orbital separation $a_0$. For a given supernova, these two parameters are practically unconstrained from the standard supernova emission alone. Our goal is to demonstrate that, if the newly born neutron star accretes from the inflated companion star, the radiative signature of the accretion power can in principle be used to constrain $M_*$ and $a_0$. However, as we will show later, it turns out that only $a_0$ is strongly constrained.

In the following, we take a logarithmic grid in both dimensions: 
\begin{equation}
    M_*/M_\odot = 1.3, 2.0, 3.0, 4.5, 6.7, 10.0,
\end{equation}
and 
\begin{equation}
    a_0/R_\odot = 12, 18, 27, 40, 60.
\end{equation}
In our model, the $a_0$'s above correspond to the pressures of the shock-heated gas $P_{\rm sh} = P_{\rm ram}$ as
\begin{equation}
    \log {P_{\rm sh}\over \rm erg\,cm^{-3}} = 14.9, 14.4, 13.9, 13.3, 12.8.
\end{equation}
Our method likely becomes less accurate at smaller orbital separations $a_0 \lesssim 10R_\odot$, because (i) the mass stripping by the ejecta's impact becomes significant, (ii) the forward shock directly passes through a large fraction of the star's volume which causes our 1D approximation to break down. 

\subsection{Results}\label{sec:MESA_results}

\begin{figure}
\centering
\includegraphics[width=0.47\textwidth]{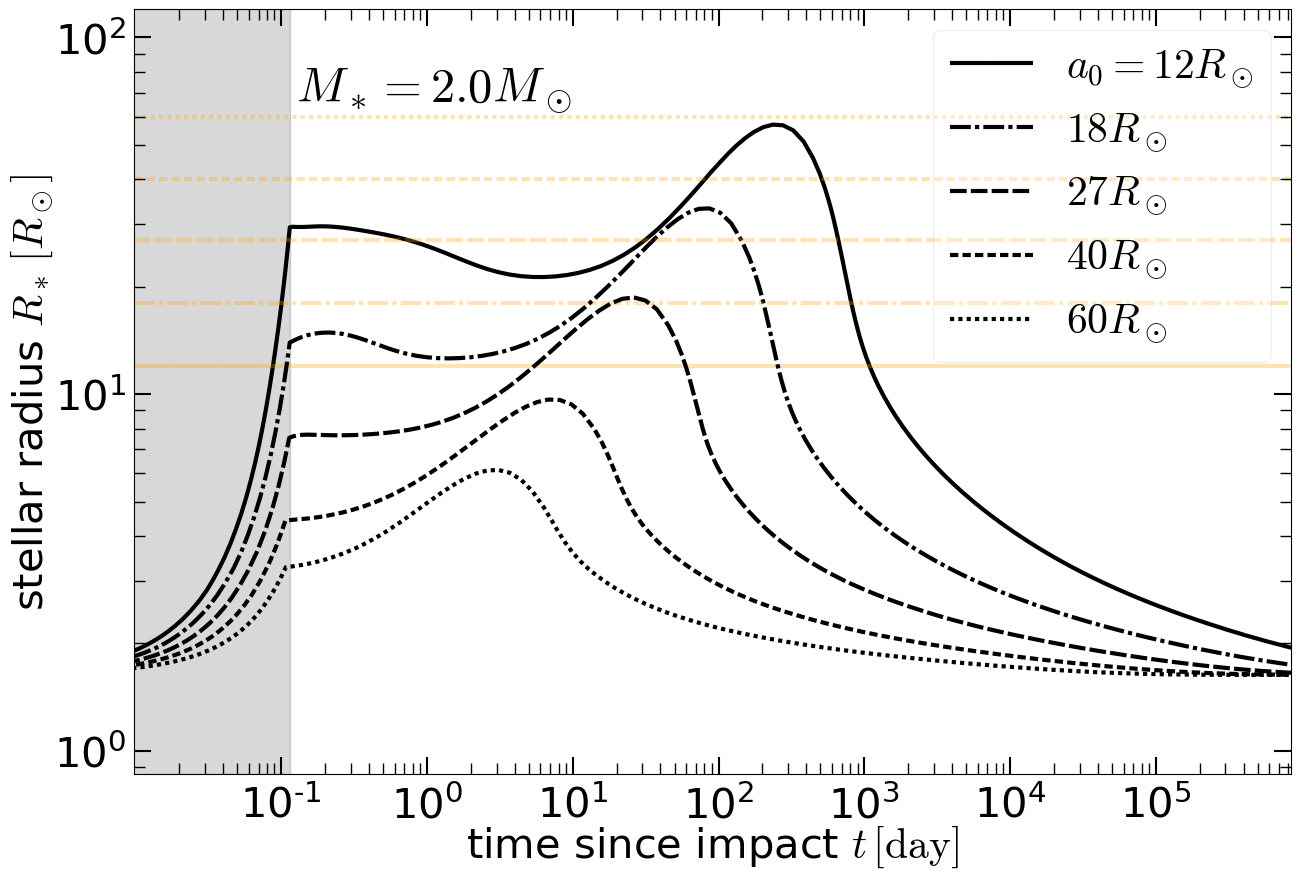}
\caption{Radius evolution (black lines) after the ejecta's impact for a $M_*=2M_\odot$ main-sequence companion star placed at different pre-supernova orbital separations $a_0$ (as shown by the horizontal orange lines). The grey region at $t<10^4\rm\, s$ marks the heat-injection episode which mimics the entropy jump due to shock-heating.
}
\label{fig:evolve_shocked_star_one_2Msun_Rstar}
\end{figure}

\begin{figure}
\centering
\includegraphics[width=0.47\textwidth]{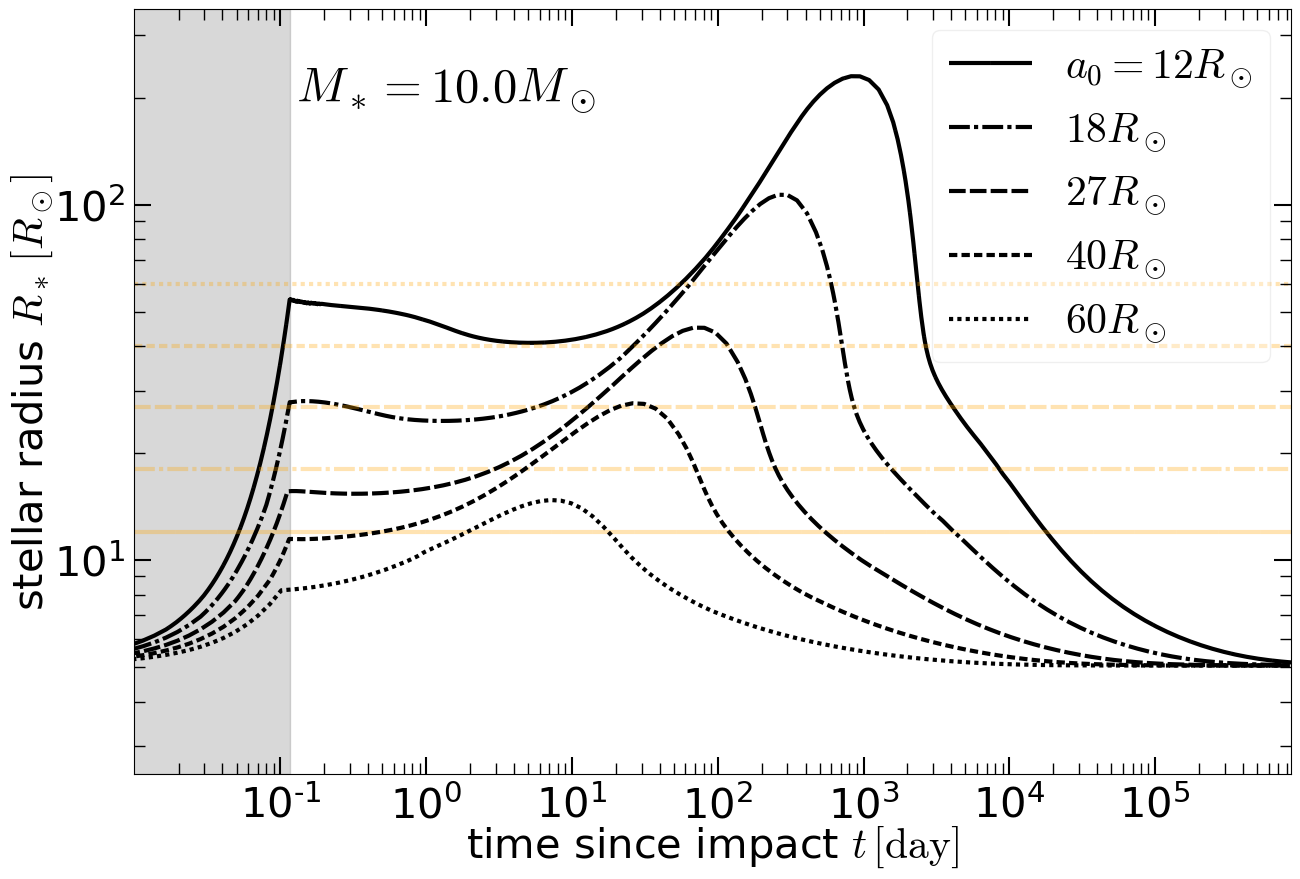}
\caption{Same as Fig. \ref{fig:evolve_shocked_star_one_2Msun_Rstar}, but for $M_*=10M_\odot$.
}
\label{fig:evolve_shocked_star_one_10Msun_Rstar}
\end{figure}

\begin{figure}
\centering
\includegraphics[width=0.47\textwidth]{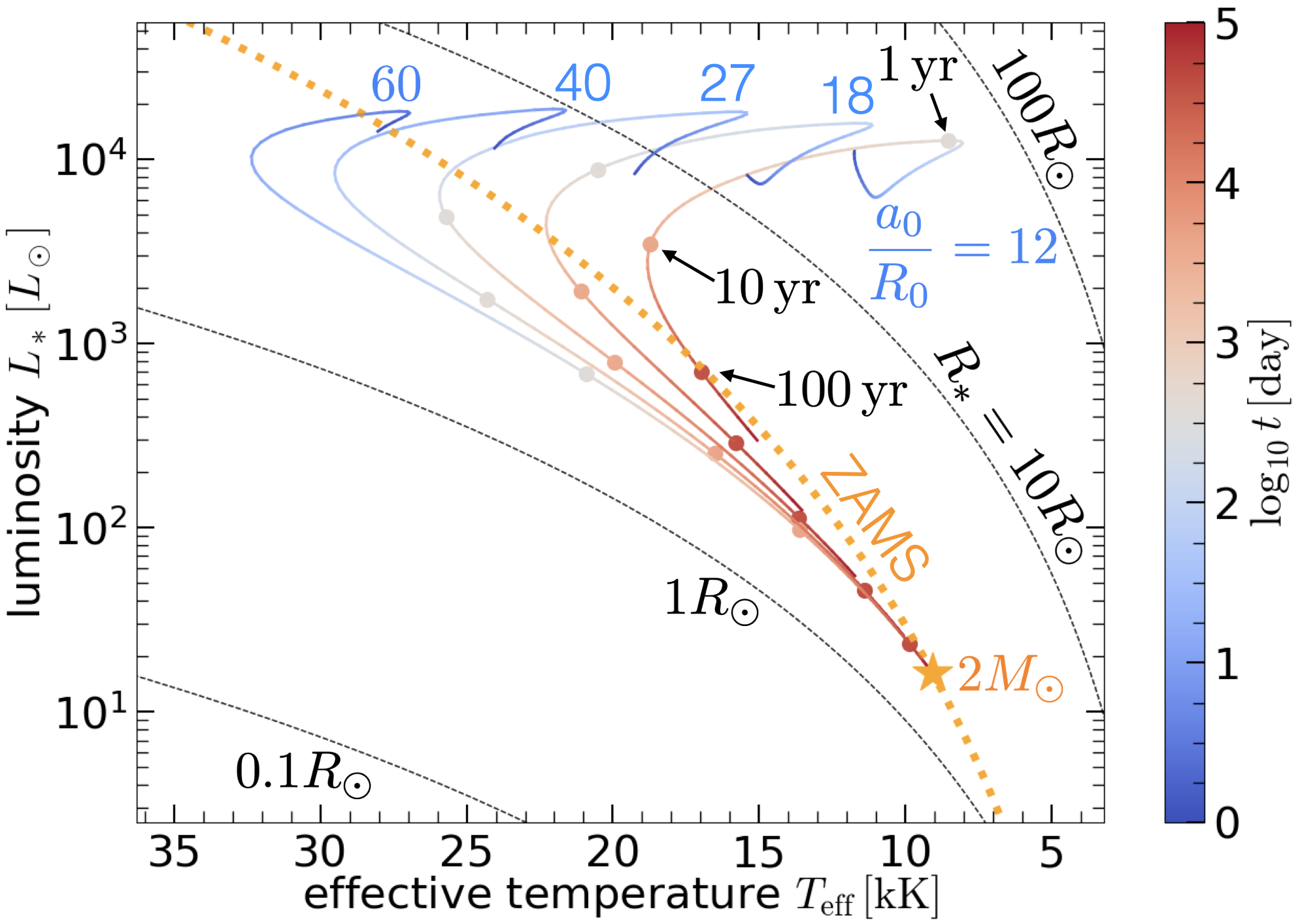}
\caption{Evolution of an ejecta-impacted $M_*=2M_\odot$ main-sequence star on the HR diagram. The colored lines correspond to different pre-supernova orbital separations $a_0$. Along each line, three filled circles indicate different times since the impact $t=1, 10, 100\rm\, yr$. The black dotted lines show different stellar radii $R_*=0.1, 1, 10, 100R_\odot$. The thick orange dotted line marks the main sequence, and the orange star represents the initial condition of the companion star. 
}
\label{fig:evolve_shocked_star_one_2Msun_HR}
\end{figure}

\begin{figure}
\centering
\includegraphics[width=0.47\textwidth]{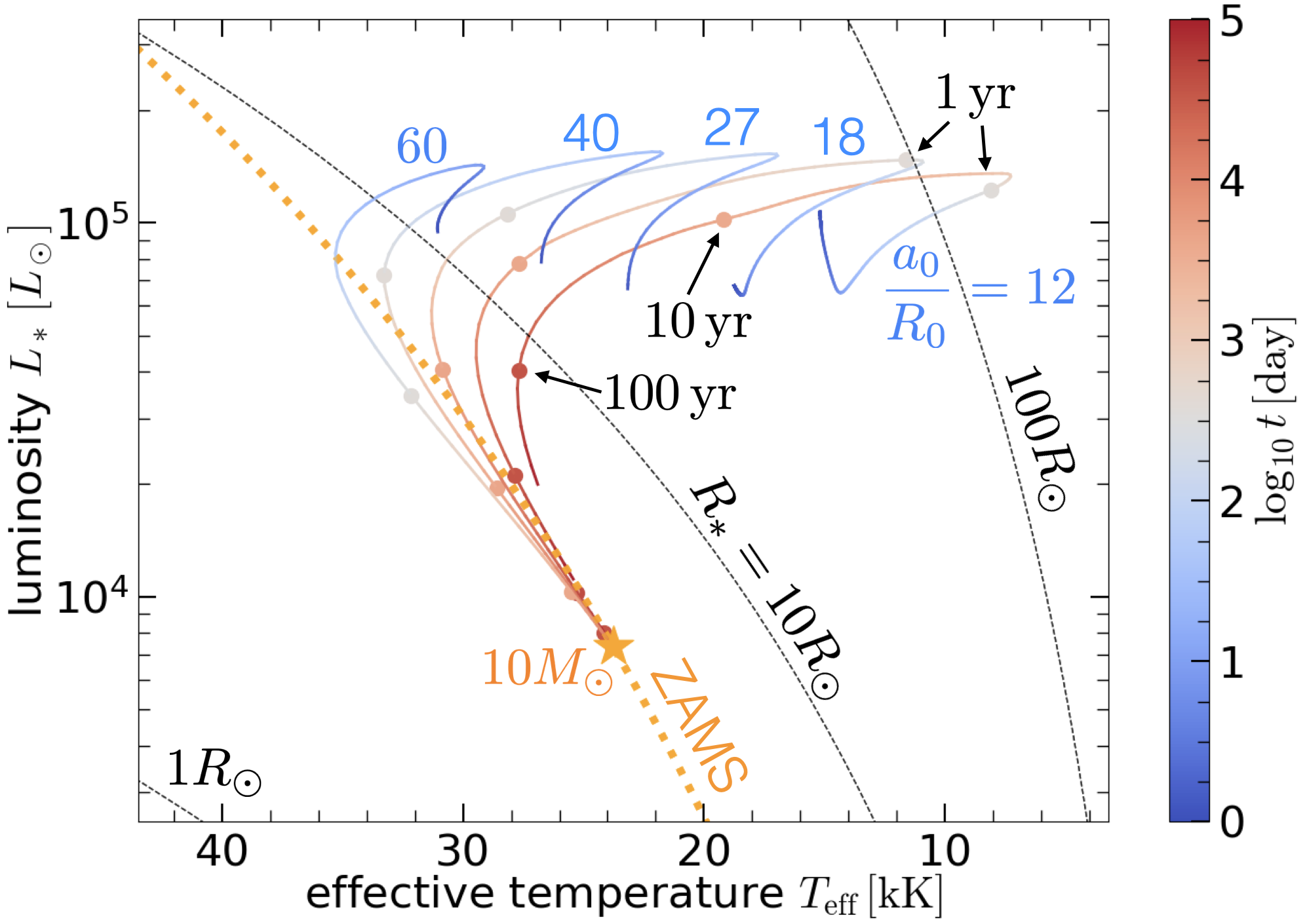}
\caption{Same as Fig. \ref{fig:evolve_shocked_star_one_2Msun_HR}, but for $M_*=10M_\odot$.
}
\label{fig:evolve_shocked_star_one_10Msun_HR}
\end{figure}

To illustrate the long-term evolution of the shock-heated star, we will focus on two representative examples with $M_*=2M_\odot$ and $10M_\odot$ and then present the metadata for other stellar masses.

The evolution of stellar radius and on the Hertzsprung-Russell (HR) diagram for the cases with $M_*=2M_\odot$ and $10M_\odot$ are shown in Figs. \ref{fig:evolve_shocked_star_one_2Msun_Rstar}--\ref{fig:evolve_shocked_star_one_10Msun_HR}. We find that the star expands to large radii: for the strongest impact considered in our grid of models, the radii grows up to $10^2R_\odot$, which is driven by hydrodynamic pressure adjustment of the outer layers. It takes up to about one year for the star to expand to the maximum radius. Subsequently, the star undergoes Kelvin-Helmholtz contraction as the shock-injected energy is radiated away. During the expansion and contraction, the star stays significantly brighter than the initial main-sequence state for an extended amount of time, ranging from decades to millennia, depending on the pre-supernova separation. These results are very similar to those found by \citet{Chen:2023aa} (see, e.g., their Fig. 7), who comprehensively discussed the observability of the shock-heated star using different optical filters.

\begin{figure}
\centering
\includegraphics[width=0.47\textwidth]{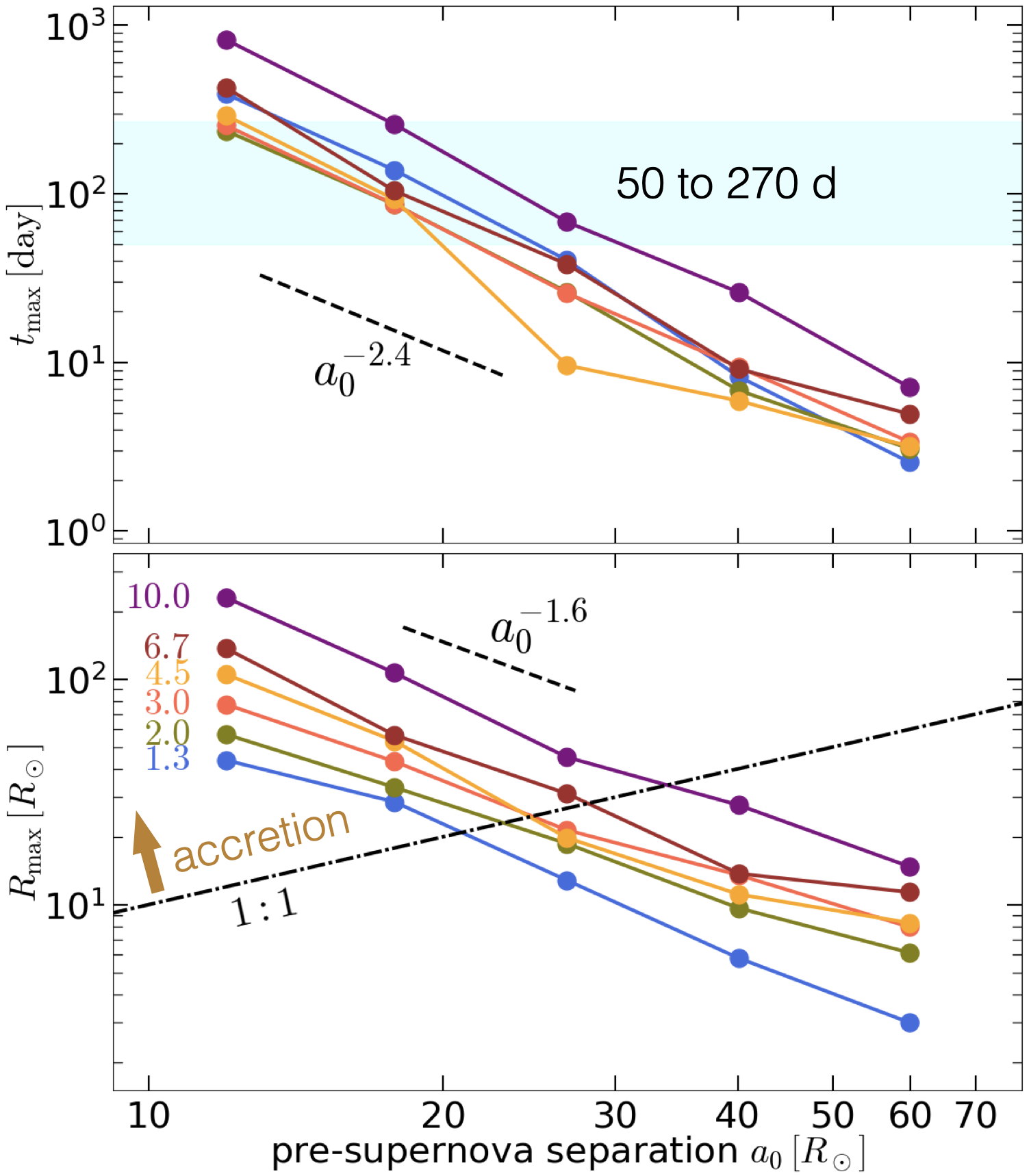}
\caption{The maximum radius $R_{\rm max}$ (lower panel) and the time to reach the maximum radius $t_{\rm max}$ (upper panel) for each of the models. In the upper panel, the cyan region marks the time between $t=50$ and $270\rm\, day$ when the inferred energy injection occurs in SN2022jli \citep{Chen:2024aa}. In the lower panel, the black dash-dotted line shows $R_{\rm max}=a_0$ above which a neutron star bound to the companion must accrete, as the post-supernova orbital pericenter radius is always less than $a_0$.
}
\label{fig:Rmax_tmax}
\end{figure}

Below, we provide a qualitative explanation of the radius expansion of the shock-heated star. Let us denote the maximum radius as $R_{\rm max}$ and the time to reach the maximum radius as $t_{\rm max}$.

Based on the dynamical timescale of the maximally inflated star, we roughly expect $t_{\rm max} \propto R_{\rm max}^{3/2}M_*^{-1/2}$. On the other hand, $t_{\rm max}$ should also be roughly equal to the Kelvin-Helmholtz cooling timescale of the inflated envelope, and this means that $t_{\rm max}\sim G M_* M_{\rm sh}/(R_{\rm max} L_{\rm Edd,*})$, where $M_{\rm sh}\propto M_* P_{\rm sh}/\bar{P}$ is the mass of the shock-heated layer (eq. \ref{eq:asymptotic_entropy_jump}). We have taken the maximum luminosity to be of the order the star's Eddington luminosity $L_{\rm Edd,*}\propto M_*$ (assuming a constant opacity). From the mass-radius relation of the unperturbed stars $R_*\propto M_*^{\approx 0.5}$ as given by our MESA models, we see that the average pressures of the stars $\bar{P}$ (eq. \ref{eq:Pbar}) are nearly independent of the masses. We then put the dynamical and cooling timescales together and find that $R_{\rm max} \propto a_0^{-6/5} M_*^{3/5}$ and $t_{\rm max}\propto a_0^{-9/5} M_*^{2/5}$, where we have used $P_{\rm sh}\propto a_0^{-3}$ for a constant explosion energy.

The above scalings are only approximate, as the detailed evolution is more complicated than the analytic estimates. Our numerical results (shown in Fig. \ref{fig:Rmax_tmax}) can be well described by the following empirical relations
\begin{equation}\label{eq:Rmax_companion_star}
    R_{\rm max} \simeq 70 R_\odot\, (a_0/10R_\odot)^{-8/5} (M_*/2M_\odot)^{1/3},
\end{equation}
and
\begin{equation}\label{eq:tmax_companion_star}
    t_{\rm max} \simeq 290\mr{\,d}\, (a_0/10R_\odot)^{-12/5}.
\end{equation}


In the following, we focus on the properties of the inflated star that are relevant for the accretion onto the neutron star under the assumption that the binary system remains bound after the supernova.


\begin{figure}
\centering
\includegraphics[width=0.47\textwidth]{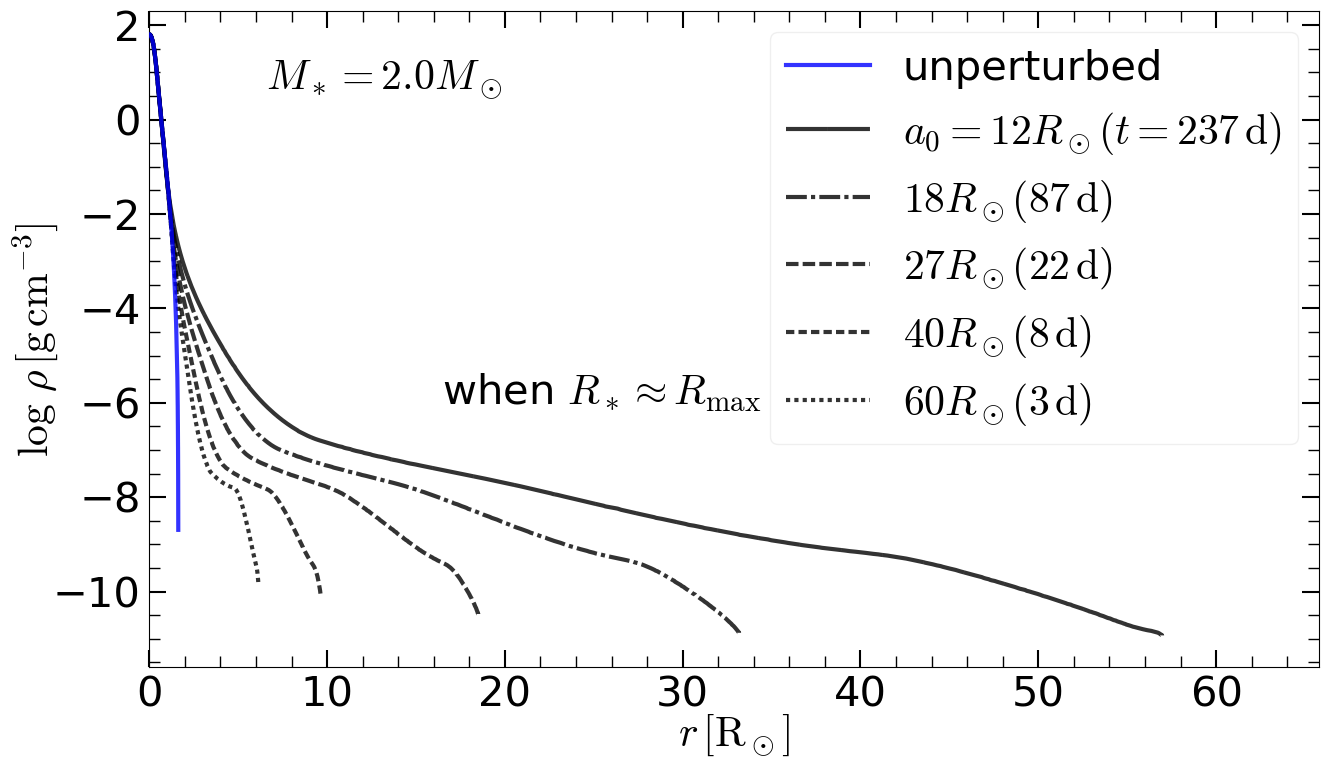}
\caption{Black lines show the density profiles of the $M_*=2M_\odot$ star near maximum radius ($R_*\approx R_{\rm max}$) for different pre-supernova orbital separations $a_0$.
The blue solid line shows the unperturbed density profile.
}
\label{fig:profile_r_rho_2Msun}
\end{figure}

Fig. \ref{fig:profile_r_rho_2Msun} shows the density profiles of the shock-inflated $M_*=2M_\odot$ star near the maximum radius $R_*\approx R_{\rm max}$ for five different pre-supernova orbital separations $a_0$. In \S \ref{sec:NS_accretion}, we will use the time-dependent density and pressure profiles to calculate the amount of gas that will be gravitationally captured by the neutron star as well as the corresponding accretion power.

For a given post-supernova orbit (which is mainly determined by the ejecta mass and the neutron star's natal kick), accretion onto the neutron star starts when the radius of the companion star $R_*$ exceeds the orbital pericenter radius $\rp$. The accretion rate is the highest when the companion star reaches the maximum radius $R_{\rm max}$. From Fig. \ref{fig:evolve_shocked_star_one_2Msun_Rstar}, we see the maximum accretion rate is achieved after a significant delay of up to about one year (depending on the pre-supernova orbital separation) after the supernova explosion. Since the radioactive decay-powered emission in a typical Type Ib/c supernova usually peaks within a few weeks after the explosion, the delay in post-supernova accretion makes it possible to form a second peak in the observed lightcurve --- this is a unique feature of our model.

In the literature, there are two other well-studied mechanisms for late-time energy injection from a compact object: (i) the magnetic dipole spin-down power from the neutron star \citep{Kasen:2010ab, Woosley:2010aa, Metzger:2015aa}, and (ii) fallback accretion of the marginally bound ejecta \cite[e.g.,][]{Dexter:2013aa, Kashiyama:2015aa, Metzger:2018aa}. These two mechanisms do not predict a second peak with a significant delay, because the energy injection is ``on'' at a very early time\footnote{Unless the magnetic dipole moment of the neutron star is enhanced in an unexpected way many weeks after the supernova \citep[see e.g.,][]{Chugai:2022aa}.} since the explosion. Another possible mechanism that can power late-time emission is the interaction between the supernova ejecta with the circum-stellar medium \citep[e.g.,][]{Chatzopoulos:2013aa, Dessart:2015aa, Khatami:2023aa}. The special case of a detached shell of circum-stellar medium can reproduce the lightcurves of some supernovae with two peaks that are clearly separated \citep{Tsuna:2023aa}, but the lightcurve and spectroscopic signatures will be different from those in our model where the heating source is located inside the supernova ejecta and the accretion power is modulated by orbital motion (see \S \ref{sec:application_SN2022jli}).


Interactions between the neutron star and companion star must occur if $R_{\rm max} > a_0$ because the post-explosion orbital pericenter radius satisfies $\rp \leq a_0$. Thus, we obtain the following \textit{sufficient} condition for accretion:
\begin{equation}\label{eq:condition_for_accretion}
    R_{\rm max}(a_0) > a_0 \ \ \Rightarrow\ \  a_0 \lesssim 20R_\odot (M_*/2M_\odot)^{0.1},
\end{equation}
which depends weakly on the companion mass.

\section{Neutron star accretion}\label{sec:NS_accretion}

In this section, we assume that the newly born neutron star remains bound to the companion star (see \S \ref{sec:orbit_constraints} for the conditions) and study the interactions between the neutron star and the shock-inflated companion star. We first estimate the mass captured by the neutron star's gravitational potential in each orbit (\S \ref{sec:Bondi_capture}) and then consider the power generated by accretion (\S \ref{sec:accretion_power}). 

\subsection{Bondi-Hoyle-Lyttleton mass capture}\label{sec:Bondi_capture}

Below, we estimate the rate of Bondi-Hoyle-Lyttleton mass capture (hereafter ``Bondi capture'') by the neutron star if its orbit goes through the companion star's inflated envelope. Let us denote the post-supernova orbital pericenter separation as $\rp$ and semimajor axis as $a$. The orbital velocity $v$ (relative velocity between the two stars) at a given separation $r$ is given by
\begin{equation}\label{eq:orbital_speed}
    v(r) = \sqrt{G(\Mns + M_*)(2/r - 1/a)},
\end{equation}
where $\Mns$ is the neutron star's mass. The critical impact parameter within which gas becomes gravitationally bound to the neutron star is given by the Bondi radius \citep{Edgar:2004aa}
\begin{equation}\label{eq:Bondi_radius}
    \rB (r) = {2G \Mns \over v^2 + \cs^2},
\end{equation}
where $\cs$ is the adiabatic sound speed of the surrounding gas, and we have ignored the companion star's gravitational potential as the Hill radius of the neutron star $r_{\rm H} = r(\Mns/M_*)^{1/3}$ (within which the neutron star's gravity dominates over that of the companion) is usually much greater than $r_{\rm B}$.

\begin{figure}
\centering
\includegraphics[width=0.4\textwidth]{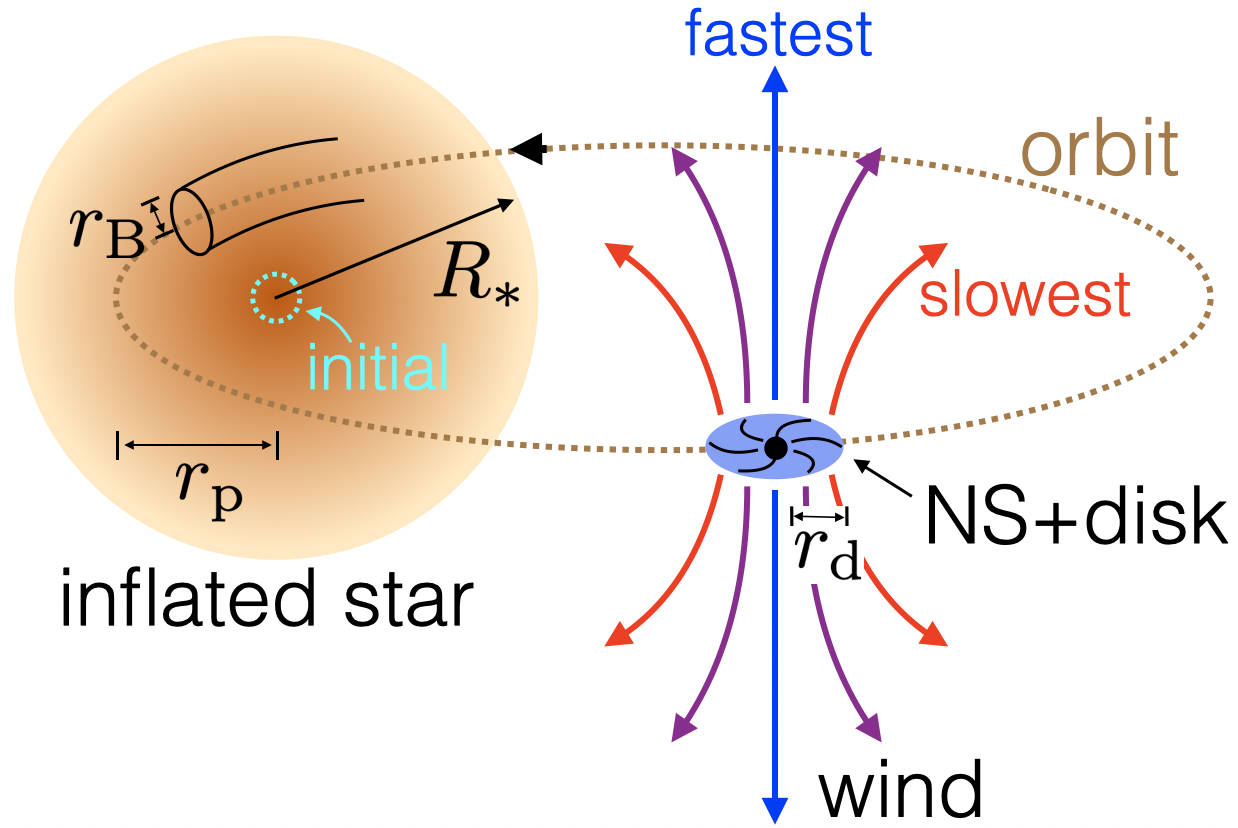}
\caption{Geometry of the post-supernova binary system.
}
\label{fig:orbit_geometry}
\end{figure}

Another important quantity is the density scale height of the stellar envelope
\begin{equation}
    H_\rho(r) \equiv |\d \ln \rho/\d r|^{-1}.
\end{equation}
This is related to the pressure scale height
\begin{equation}
    H_P(r) \equiv |\d \ln P/\d r|^{-1} \approx {P\over \rho} {r^2 \over GM_*} = {\cs^2 r^2\over \gamma GM_*},
\end{equation}
where $\gamma\approx 5/3$ is the adiabatic index of the gas. We assume quasi-hydrostatic equilibrium in the envelope, and we have taken $m(r)\approx M_*$ as the envelope only contains a small fraction of the star's mass. These two scale heights are related to each other by
\begin{equation}
    {H_\rho\over H_P} = {\d \ln P\over \d \ln \rho} \equiv \Gamma,
\end{equation}
where $\Gamma$ is a factor of order unity that depends on the pressure and density profiles of the envelope. In the following, we adopt\footnote{Unfortunately, ${\d \ln P/\d \ln \rho}$ is a derived quantity in MESA that is only correct to the first order of the radial grid spacing, so it has significant numerical fluctuations near the stellar surface. The pressure scale height $H_P$ is quantity that is correct to a higher order and directly saved in MESA output files, our model relies on $H_P(r)$ and a constant, empirical $\Gamma$. }
$\Gamma\simeq 1.5$ and hence
\begin{equation}\label{eq:density_scale_height}
    H_\rho \simeq 1.5 H_P,
\end{equation}
where the factor of 1.5 comes from our numerical tests and is appropriate for regions deep inside the inflated envelope at radii of the order $10R_\odot$ (where efficient Bondi capture onto the neutron star occurs). 


The ratio between the density scale-height and the Bondi radius is roughly given by
\begin{equation}
    {H_\rho \over \rB } \simeq {\cs^2\over v^2} {2\Gamma (1 + \cs^2/v^2)\over \gamma} {(\Mns + M_*)^2\over \Mns M_*},
\end{equation}
where we have used a rough estimate of $v\simeq \sqrt{2G(\Mns + M_*)/r}$ as the typical orbital speed of the neutron star near the pericenter (where most mass is captured). Since the neutron star is moving supersonically through the envelope ($\cs \ll v$), the density scale-height $H_\rho$ is typically less than the Bondi radius $\rB$. This means that density profile across the ``Bondi cylinder'' of radius $\rB$ is inhomogeneous. The side closer to the stellar center has higher density than the farther side. Therefore, only the fluid elements with impact parameters $\lesssim H_\rho$ will undergo strong shock interactions and could further fall into the gravitational potential of the neutron star. The trajectories of the fluid elements with impact parameters in between $H_\rho$ and $\rB$ are significantly deflected by the neutron star's gravity, but they largely maintain hyperbolic orbits and stay unbound to the neutron star during the encounter. Such a system has been numerically studied by \citet{MacLeod:2015aa, MacLeod:2017aa, De:2020aa} and their results are consistent with the picture where the critical cylindrical radius within which the gas gets physically captured by the neutron star is given by $H_\rho$ instead of $\rB$, if $H_\rho < \rB$.




Since most of the mass is captured near the pericenter $\rp$ of the orbit, we adopt the following estimate for the captured mass \textit{per orbit}
\begin{equation}\label{eq:captured_mass_per_orbit}
    \Delta M_{\rm cap} \simeq \pi^2 \rp \rho(\rp) \min(H_\rho^2, \rB^2)|_{\rp},
\end{equation}
where we have taken a path length of $\pi \rp$ near the pericenter. We emphasize that eq. (\ref{eq:captured_mass_per_orbit}) does not take into account the gas accumulation near the neutron star --- an existing accretion disk (from earlier gas capture) may increase the effective cross-section for mass capture. We also ignore the captured mass at orbital separations much greater than $\rp$. For these reasons, our estimated $\Delta M_{\rm cap}$ is likely a lower limit to the true values. 

Since the neutron star moves super-sonically through the envelope, the perturbation to the envelope structure by the neutron star's gravitational acceleration will not change our results substantially. However, we expect the modification of the envelope structure by the feedback from the accretion power to be important, for the following reason \citep[see also][]{Hober:2022aa}. Suppose the accretion power $L=10^{42}L_{42}\rm\, erg\,s^{-1}$ is carried by an outflow with speed $v_{\rm w}$, this produces a ram pressure at radius $r$ from the neutron star
\begin{equation}\label{eq:disk_wind_pressure}
    P_{\rm w} = {L\over 4\pi r^2 v_{\rm w}} \simeq 5.5\times 10^{7}\mr{erg\over cm^{3}} L_{42} {0.1c\over v_{\rm w}} \lrb{10R_\odot\over r}^{2}.
\end{equation}
This pressure $P_{\rm w}$ is comparable to the ram pressure of the Bondi-captured gas in the neutron star's frame
\begin{equation}\label{eq:ram_pres_accretion_flow}
    \rho v^2 \simeq 10^8\mr{erg\over cm^{3}} \rho_{-7} (v/300\rm\, km\,s^{-1})^2,
\end{equation}
where $\rho = 10^{-7}\rho_{-7}\rm\, g\,cm^{-3}$ is the density of the envelope near the orbital pericenter radius $\rp\simeq 10R_\odot$ (see Fig. \ref{fig:profile_r_rho_2Msun}) and $v$ is the relative velocity between the gas and the neutron star.



Nevertheless, the above issues of our crude model can only be addressed by detailed numerical simulations of the interactions between a neutron star and the inflated stellar envelope, which is left for future works.


\begin{figure}
\centering
\includegraphics[width=0.47\textwidth]{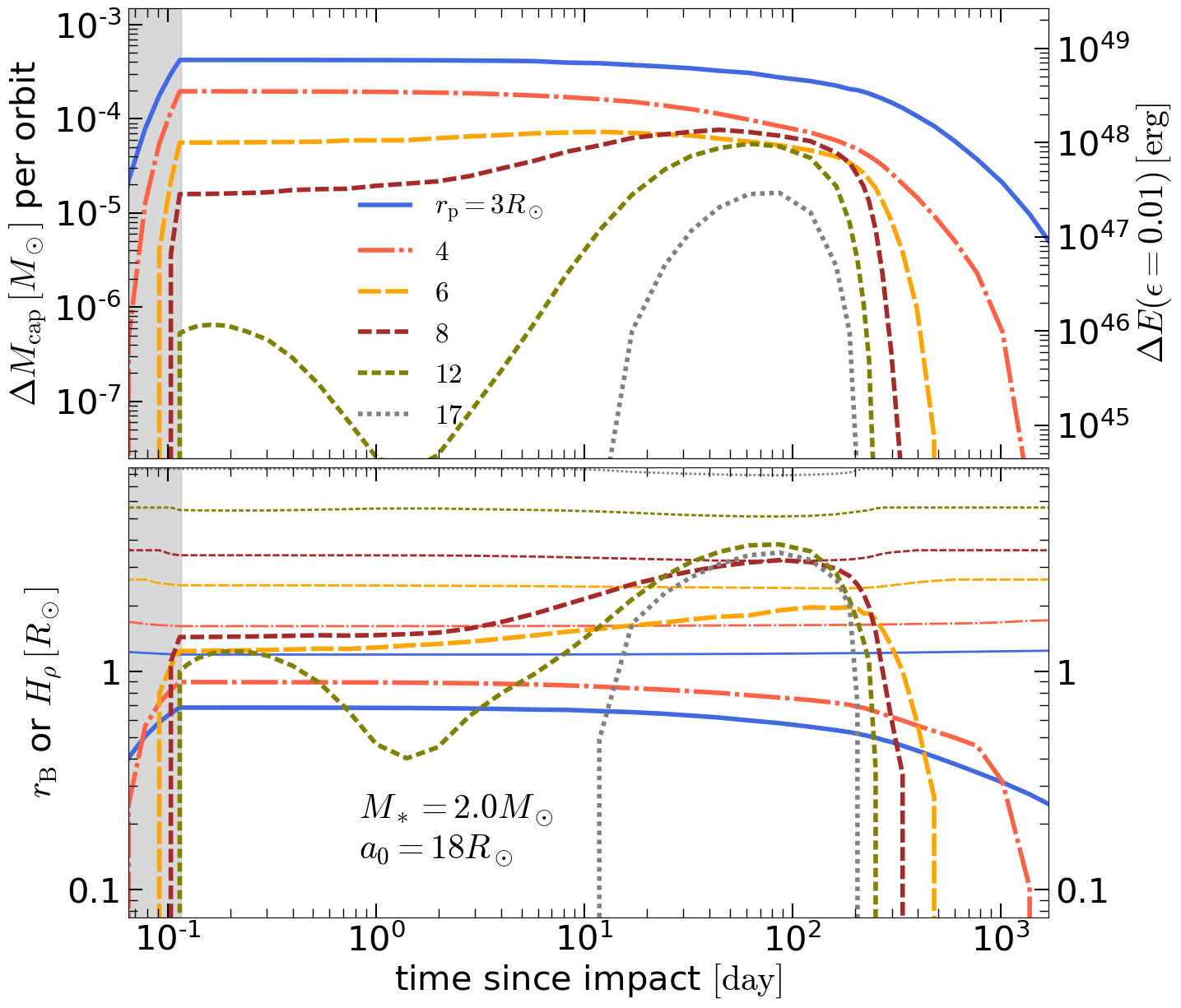}
\caption{\textit{Upper panel}: Captured mass per orbit $\Delta M_{\rm cap}$ (eq. \ref{eq:captured_mass_per_orbit}) and the energy release $\Delta E=\eps \Delta M_{\rm cap}c^2$ for different pericenter radii $\rp/R_\odot$ and a fiducial accretion efficiency $\eps=0.01$. Since $\Delta M_{\rm cap}$ only depends weakly on the post-supernova semimajor axis $a$ (as long as the orbit is at least mildly eccentric), we fix $a=50R_\odot$ in all cases. The companion is a $M_*=2.0M_\odot$ main-sequence star at pre-supernova orbital separation of $a_0=18R_\odot$. \textit{Lower panel}: The density scale-height $H_\rho$ (thick lines, eq. \ref{eq:density_scale_height}) and Bondi radius $r_{\rm B}$ (thin lines, eq. \ref{eq:Bondi_radius}), both evaluated at the post-supernova pericenter $r=\rp$. We find $H_\rho\lesssim \rB$ in most cases. In both panels, the gray region at $t<10^4\rm\, s$ marks the heat-injection episode.
}
\label{fig:Mcapture2.0_a18}
\end{figure}

The time evolution for the captured mass per orbit $\Delta M_{\rm cap}$ for a $M_*=2M_\odot$ companion star placed at a pre-supernova orbital separation of $a_0=18R_\odot$ (corresponding to $P_{\rm sh}=2.4\times10^{14}\rm\, erg\,cm^{-3}$) is shown in Fig. \ref{fig:Mcapture2.0_a18}. Note that the time axis should be understood as the pericenter passage time for a hypothetical orbit --- the binary orbital motion is not explicitly considered in this calculation.

For the post-supernova orbit (which depends on the unknown natal kick on the neutron star, see \S \ref{sec:orbit_constraints}), we consider a wide range of pericenter radii $3 \leq \rp \leq 17R_\odot$ and fix the semimajor axis to be $a=50R_\odot$ (the results depend weakly on $a$ as long as $2/\rp \gg 1/a$ in eq. \ref{eq:orbital_speed}). Note that, for any natal kick, the post-supernova pericenter radius is physically constrained by $\rp \leq a_0$. On the other hand, we do not consider very small pericenter radii, because the companion star will likely be tidally disrupted by the neutron star \citep{Perets:2016aa, Kremer:2019aa, Kremer:2022aa} and our Bondi-capture picture breaks down. The observational consequences of such tidal disruptions are calculated in detail by \citet{tsuna25_TDEs}.

We find that, for $a_0=18R_\odot$, when the companion star expands to near its maximum radius, the neutron star typically captures a mass of $\Delta M_{\rm cap}\sim 10^{-4} M_\odot$ per orbit. For an accretion efficiency $\eps\sim 10^{-2}$ (see \S \ref{sec:accretion_power} and eq. \ref{eq:max_acc_efficiency}), then accretion onto the neutron star can produce $10^{48}\rm\, erg$ of energy (right axis in Fig. \ref{fig:Mcapture2.0_a18}).

For smaller post-supernova pericenter radii $\rp\lesssim 8R_\odot$, the mass capture starts immediately after the ejecta's impact (i.e., in the first post-supernova orbit). This is because the radius of the shock-heated star $R_*$ exceeds $\rp$ by the end of the heating episode (see Fig. \ref{fig:evolve_shocked_star_one_2Msun_Rstar}). The mass capture persists until $\gtrsim1\rm\, yr$ after the supernova --- the time frame is longer for smaller $\rp$, and then the mass capture shuts off due to Kelvin-Helmholtz contraction of the shock-heated outer envelope. For larger pericenter radii $\rp \gtrsim 12R_\odot$ (green dashed lines), the strongest mass capture occurs after a substantial delay of the order 10 days to one month, and the mass capture ends abruptly after about 200 days. Later in \S \ref{sec:application_SN2022jli}, we will discuss the observational implications of such a delayed onset of neutron star accretion --- such systems will produce unique, double-peaked supernova lightcurves.

\begin{figure}
\centering
\includegraphics[width=0.47\textwidth]{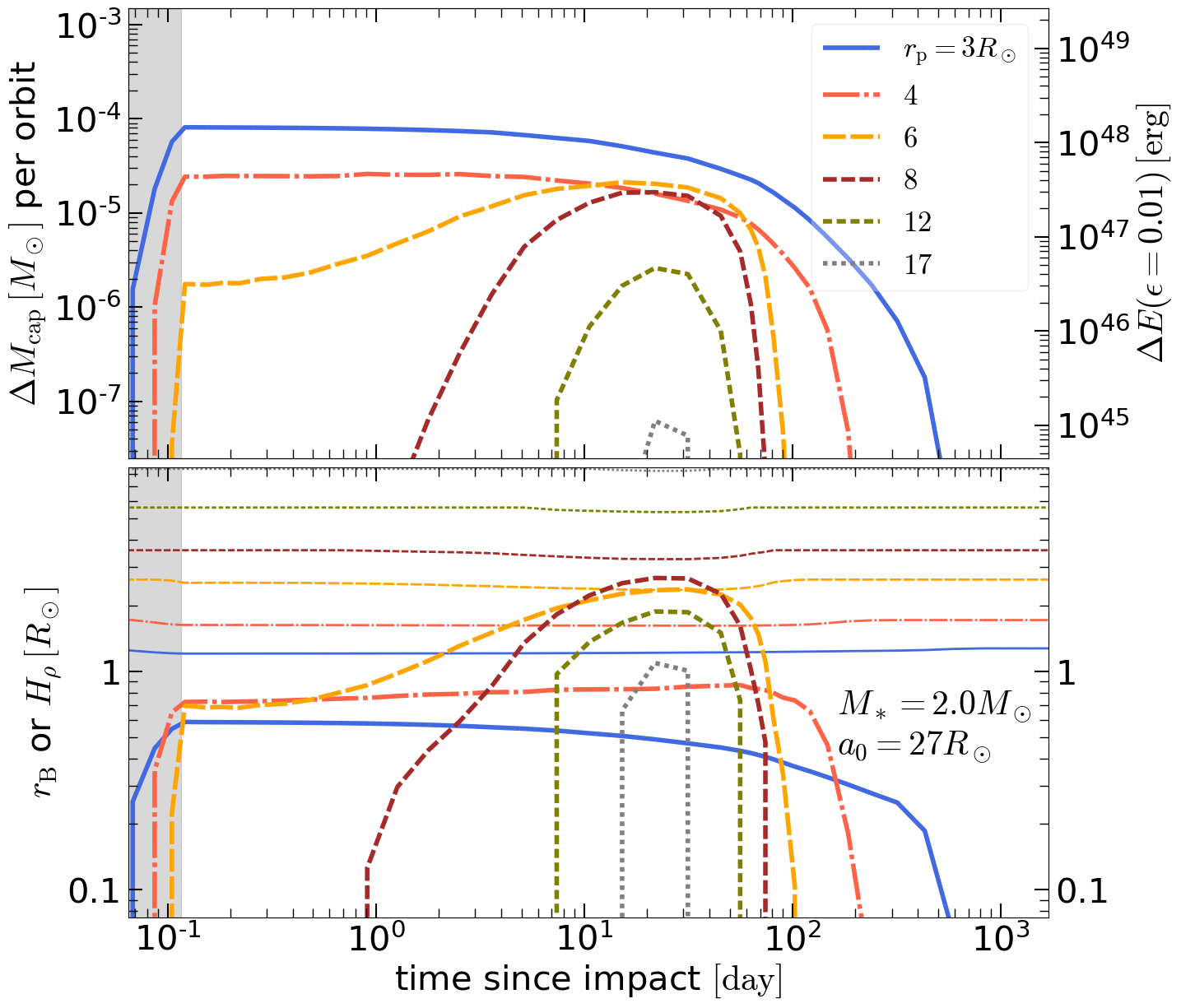}
\caption{The same as Fig. \ref{fig:Mcapture2.0_a18} but for pre-supernova orbital separation of $a_0=27R_\odot$.
}
\label{fig:Mcapture2.0_a27}
\end{figure}

In Fig.~\ref{fig:Mcapture2.0_a27}, we show another case for a larger pre-supernova orbital separation of $a_0=27R_\odot$ (corresponding to $P_{\rm sh}=7.4\times10^{13}\rm\, erg\,cm^{-3}$). The captured mass per orbit is significantly less than that for $a_0=18R_\odot$, because the envelope is less inflated under a weaker ram pressure. However, the qualitative results are similar: for smaller $\rp\lesssim 6R_\odot$, the mass capture is stronger and starts promptly; whereas for larger $\rp\gtrsim 8R_\odot$, the mass capture is weaker and reaches a peak after some delay.



In the following, we assume $H_\rho(\rp) \lesssim \rB(\rp)$ (as shown by the lower panels of Figs. \ref{fig:Mcapture2.0_a18} and \ref{fig:Mcapture2.0_a27}) and calculate the specific angular momentum of the captured gas. When the neutron star reaches the pericenter, the gas located within a cylindrical radius $\tilde{r}\lesssim  H_\rho(\rp)$ in the plane perpendicular to the neutron star's velocity $v$ will be efficiently captured. Let us denote the azimuthal angle in the perpendicular plane as $\tilde{\phi}$, with $\tilde{\phi}=0$ corresponding to the radial direction outwards from the center of the companion star. To the linear order, the gas density in the perpendicular plane at coordinate $(\tilde{r}, \tilde{\phi})$ is given by
\begin{equation}
    \rho(\tilde{r}, \tilde{\phi}) \approx \rho_0 - {\rho_0\over H_\rho} \tilde{r}\cos\tilde{\phi},
\end{equation}
where $\rho_0 = \rho(\rp)$ is the stellar density at $r=\rp$ and $H_\rho = H_\rho(\rp)$.
The mass capture rate is given by
\begin{equation}
    \dot{M}_{\rm d}(\rp) \simeq \int_0^{2\pi}\d\tilde{\phi} \int_0^{H_\rho} \d \tilde{r} \, \tilde{r} \rho(\tilde{r}, \tilde{\phi}) v_{\rm p},
\end{equation}
and the angular momentum capture rate is
\begin{equation}
    \dot{L}_{\rm d}(\rp) \simeq \int_0^{2\pi}\d\tilde{\phi} \int_0^{H_\rho} \d \tilde{r}\, \tilde{r} \rho(\tilde{r}, \tilde{\phi}) v_{\rm p} \, \tilde{r} v_{\rm p} \cos\tilde{\phi},
\end{equation}
where $v_{\rm p} = v(\rp)$ is the velocity at the pericenter and the extra factor of $\cos\tilde{\phi}$ in $\dot{L}_{\rm d}(\rp)$ comes from the fact that only the angular momentum component perpendicular to the post-supernova orbital plane is non-zero. Thus, we obtain the specific angular momentum for the accretion disk around the neutron star
\begin{equation}
\begin{split}
    \ell_{\rm d} &\simeq {\dot{L}_{\rm d}(\rp) \over \dot{M}_{\rm d}(\rp)} \simeq {1\over 4} H_\rho(\rp) v(\rp).
\end{split}
\end{equation}
Since $v(\rp)\approx \sqrt{2G(\Mns + M_*)/\rp}$, we then obtain the circularization radius of the accretion disk
\begin{equation}
    r_{\rm d,0} = {\ell_{\rm d}^2\over G\Mns} \simeq {H_\rho^2\over 8\rp} \lrb{1 + {M_*\over \Mns}}.
\end{equation}
For typical values of $H_\rho \sim 3R_\odot$, $\rp \sim 10R_\odot$, and $M_*\sim 3M_\odot$, we obtain
\begin{equation}
    r_{\rm d,0} \sim 0.3R_\odot.
\end{equation}


As the disk will viscously expand over time, its size $\rd$ is roughly given by the condition that the viscous timescale,
\begin{equation}\label{eq:viscous_time}
    t_{\rm vis} = {1\over \alpha h^{2}\Omega_{\rm K}(\rd)} \simeq 0.3\mr{\,d}\, {(\rd/0.3 R_\odot)^{3/2}} {\alpha h^2 \over 0.01},
\end{equation}
is comparable to the mass-capture timescale of $\sqrt{\rp^3/G(\Mns+M_*)}$, so we obtain the disk radius
\begin{equation}\label{eq:disk_radius}
\begin{split}
    \rd &\simeq \rp {(\alpha h^2)^{2/3}\over (1 + M_*/\Mns)^{1/3}}\\
    &\simeq 0.3R_\odot {\rp\over 10R_\odot} \lrb{\alpha h^2\over 0.01}^{2/3} \lrb{3\over 1 + M_*/\Mns}^{1/3},
\end{split}
\end{equation}
where $\alpha$ is the dimensionless viscosity parameter \citep{Shakura:1973aa}, $h$ is the ratio between the vertical scale height and the disk radius $\rd$, and $\Omega_{\rm K}(\rd) = \sqrt{G\Mns/\rd^3}$ is the Keplerian angular frequency. Hereafter, we take $\rd \simeq 0.3 R_\odot$ to be our fiducial value for the disk size.

Then, the characteristic accretion rate in the outer disk is given by
\begin{equation}\label{eq:acc_rate_outer_disk}
    \dot{M}_{\rm d} \simeq {\Delta M_{\rm cap}\over t_{\rm vis}} \simeq 0.14 {M_\odot\over \rm yr}\, {\Delta M_{\rm cap}\over 10^{-4}M_\odot} {\alpha h^{2}/0.01\over  (\rd/0.3R_\odot)^{3/2}}.
\end{equation}
Below, we show that such a high accretion rate can potentially power very bright emission.

\subsection{Accretion onto magnetized neutron star}\label{sec:accretion_power}


In this section, we consider the power generated by the accretion flow around the neutron star.

The Eddington luminosity of the neutron star is given by $L_{\rm Edd} =4\pi G\Mns c/\kappa_{\rm T}$, where $\kappa_{\rm T}\approx 0.34\rm\,cm^2\,g^{-1}$ is the electron scattering opacity appropriate for the accreting gas. The key property of the accretion flow is the \textit{locally} super-Eddington accretion rate $GM_{\rm ns}\dot{M}(r)/r\gg L_{\rm Edd}$ at relevant radii $r$ between the neutron star radius $R_{\rm ns}$ and the disk outer radius $r_{\rm d}$ (eq. \ref{eq:disk_radius}). Thermally driven outflow causes the accretion rate to drop towards smaller radii, and the self-similar solution is given by \citep{Narayan:1995aa, Blandford:1999aa}
\begin{equation}
    \dot{M}(r) = \dot{M}_{\rm d} (r/\rd)^p,
\end{equation}
where $\dot{M}_{\rm d}$ is the mass accretion rate near the disk outer radius and $p$ is the power-law index for the radial scaling of the accretion rate. Following the results from numerical simulations \citep{Stone:1999aa, Stone:2001aa, Yuan:2012aa, Yuan:2015aa, White:2020aa, Cho:2023aa, Guo:2024aa}, we take $p= 0.5$ in this paper as a representative value. For completeness, we take $\rd$ in the above power-law scaling to be the spherization radius $r_{\rm sph} = GM_{\rm ns}\dot{M}_{\rm d}/L_{\rm Edd}$ \citep{Begelman:1979aa}, when $r_{\rm sph}<\rd$.


The accretion flow is truncated at the Alfv\'en radius $\rA$ where the magnetospheric pressure overwhelms the gas pressure. For quasi-spherical accretion, this critical radius is given by \citep{Davidson:1973aa, Ghosh:1979aa, Frank:2002aa}
\begin{equation}
    r_{\rm A} \simeq \lrb{\muB^2 \over 2\dot{M}(r_{\rm A}) \sqrt{2GM_{\rm ns}}}^{2/7},
\end{equation}
or 
\begin{equation}
    r_{\rm A} \simeq \lrb{\muB^2 r_{\rm d}^p \over 2\dot{M}_{\rm d} \sqrt{2GM_{\rm ns}}}^{2\over 7+2p},
\end{equation}
where $\mu_{\rm B} = B_{\rm ns} R_{\rm ns}^3$ is the magnetic dipole moment of the neutron star and $B_{\rm ns}$ is the surface dipolar B-field strength at the magnetic equator.
The corotation radius is where the Keplerian frequency is equal to the spin frequency and given by
\begin{equation}
    r_{\rm co} = \lrb{GM_{\rm ns}\over \Omega^2}^{1/3},
\end{equation}
where $\Omega=2\pi/P$ is the spin angular frequency.
The light cylinder radius is
\begin{equation}
    r_{\rm lc} = c/\Omega.
\end{equation}
The interplay among these three radii $\rA$, $r_{\rm co}$, and $r_{\rm lc}$ determines the total power of the neutron star accretion flow system, which consists of four components:
\begin{itemize}
    \item[(i)] The electromagnetic (Poynting) power $L_{\rm m}$ from the open magnetic field lines, accounting for the disk-induced opening of fields \citep{Parfrey:2016aa, Metzger:2018aa},
\begin{equation}
    L_{\rm m} = {\muB^2\Omega^4\over c^3} \max\lrsb{1, \lrb{r_{\rm lc}/r_{\rm A}}^2}.
\end{equation}
    \item[(ii)] The power from accretion onto the neutron star's surface
    \begin{equation}
        L_{\rm ns} = \min\lrb{L_{\rm ns,max}, {GM_{\rm ns} \dot{M}_{\rm ns} \over R_{\rm ns}}},
    \end{equation}
    where $L_{\rm ns,max} = 3.5\times 10^{39} (B_{\rm ns}/10^{12}\mr{\,G})^{3/4}\rm\, erg\,s^{-1}$ is the maximum luminosity from the accretion column onto the neutron star \citep{Basko:1976aa, Mushtukov:2015aa}, and the surface accretion rate is given by $\dot{M}_{\rm ns} = \dot{M}(\rA)$ if $r_{\rm A} < \min(r_{\rm lc}, r_{\rm co})$ (in the accretor state) and $\dot{M}_{\rm ns}\approx 0$ otherwise (since very little mass will land on the neutron star). Note that when the neutron star accretes at extremely high rates, a large fraction of the accretion power is lost due to neutrino emission, which causes the radiative power to be limited to $L_{\rm ns,max}$ \citep{Basko:1976aa}.
    \item[(iii)] If $r_{\rm co} < r_{\rm A} < r_{\rm lc}$, the system is in the propeller state, so the gas near the Alfv\'en radius is forced to corotate at the spin angular frequency and this forms a wind whose power is given by
    \begin{equation}
        L_{\rm w} = {(\Omega\, r_{\rm A})^2 \dot{M}(\rA) }.
    \end{equation}
    When the system is not in the propeller regime, we take $L_{\rm w}=0$. 
    \item[(iv)] The disk accretion power near the Alfv\'en radius is given by $L_{\rm d} = {G\Mns \dot{M}(\rA)/\rA}$, where $\dot{M}(\rA) = \dot{M}_{\rm d}(r_{\rm A}/r_{\rm d})^p$ is the accretion rate near $\rA$.
\end{itemize}

Thus, the total power from the accretion system in different states are given by
\begin{equation}
    L \simeq
    \begin{cases}
        L_{\rm m}, \ &\mbox{ if } r_{\rm A} > r_{\rm lc} \mbox{ (pulsar)},\\
        L_{\rm m} + L_{\rm w} + L_{\rm d}, \ &\mbox{ if } r_{\rm co} < r_{\rm A} < r_{\rm lc} \mbox{ (propeller)},\\
        L_{\rm m} + L_{\rm ns} + L_{\rm d},\ &\mbox{ if } r_{\rm A} < \min(r_{\rm lc}, r_{\rm co}) \mbox{ (accretor)}.
    \end{cases}
\end{equation}

An important outcome is that the outflows have a wide range of velocities from $\sim\sqrt{G\Mns/\rd}$ all the way up to near the speed of light (when $\rA$ is close to $\rlc$ but above $\rco$). The velocity stratification leads to the formation of internal shocks when the faster components launched in a later orbit collide with the slower components launched in an earlier orbit. The emission from such internal shocks will be discussed in \S \ref{sec:shock_emission}.

The neutron star's spin angular frequency $\Omega$ is affected by the spin-up and spin-down torques according to
\begin{equation}\label{eq:spin_torques}
    I_{\rm ns} \dot{\Omega} \simeq
    \begin{cases}
        -{L_{\rm m}\over \Omega}, \ &\mbox{ if } r_{\rm A} > r_{\rm lc},\\
        -{L_{\rm m}\over \Omega} + \dot{M}(\rA) \rA^2 \lrsb{\Omega_{\rm K}(\rA) - \Omega}, \ &\mbox{ else},
    \end{cases}
\end{equation}
where $I_{\rm ns}$ is the neutron star's moment of inertia, and $\Omega_{\rm K}(r_{\rm A}) = \sqrt{G M_{\rm ns}/r_{\rm A}^3}$ is the Keplerian angular frequency at the Alfv\'en radius. The torques follow the prescriptions by \citet{Eksi:2005aa, Piro:2011aa}, although there are theoretical uncertainties about the detailed magnetosphere-disk interactions near the Alfv\'en radius \citep[e.g.,][]{Wang:1995aa, Lovelace:1995aa, Miller:1997aa, Romanova:2004aa, Parfrey:2017aa}.

When the electromagnetic torque $-L_{\rm m}/\Omega$ is negligible compared to other components, the equilibrium spin frequency $\tilde{\Omega}_{\rm eq}$ is given by $r_{\rm A} = r_{\rm co}(\tilde{\Omega}_{\rm eq})$, or
\begin{equation}
    \tilde{\Omega}_{\rm eq} = {2\pi\over \tilde{P}_{\rm eq}} = \lrb{GM_{\rm ns}}^{5+p\over 7+2p} \lrb{\dot{M}_{\rm d} \over \muB^2 r_{\rm d}^p}^{3\over 7+2p}.
\end{equation}
In reality, the true equilibrium spin frequency ${\Omega}_{\rm eq}$ as given by $\dot{\Omega}=0$ is slightly different from $\tilde{\Omega}_{\rm eq}$. Nevertheless, $\tilde{\Omega}_{\rm eq}$ is important as it separates the propeller state from the accretor state. Another critical spin frequency $\Omega_{\rm cr}$ is given by $r_{\rm A}=r_{\rm lc}$, which sets the boundary between the pulsar state ($\rA > \rlc$) and the propeller/accretor states ($\rA < \rlc$), and we have
\begin{equation}\label{eq:Omega_cr}
    \Omega_{\rm cr} = 2\pi/P_{\rm cr} = c/r_{\rm A}.
\end{equation}

\begin{figure}
\centering
\includegraphics[width=0.48\textwidth]{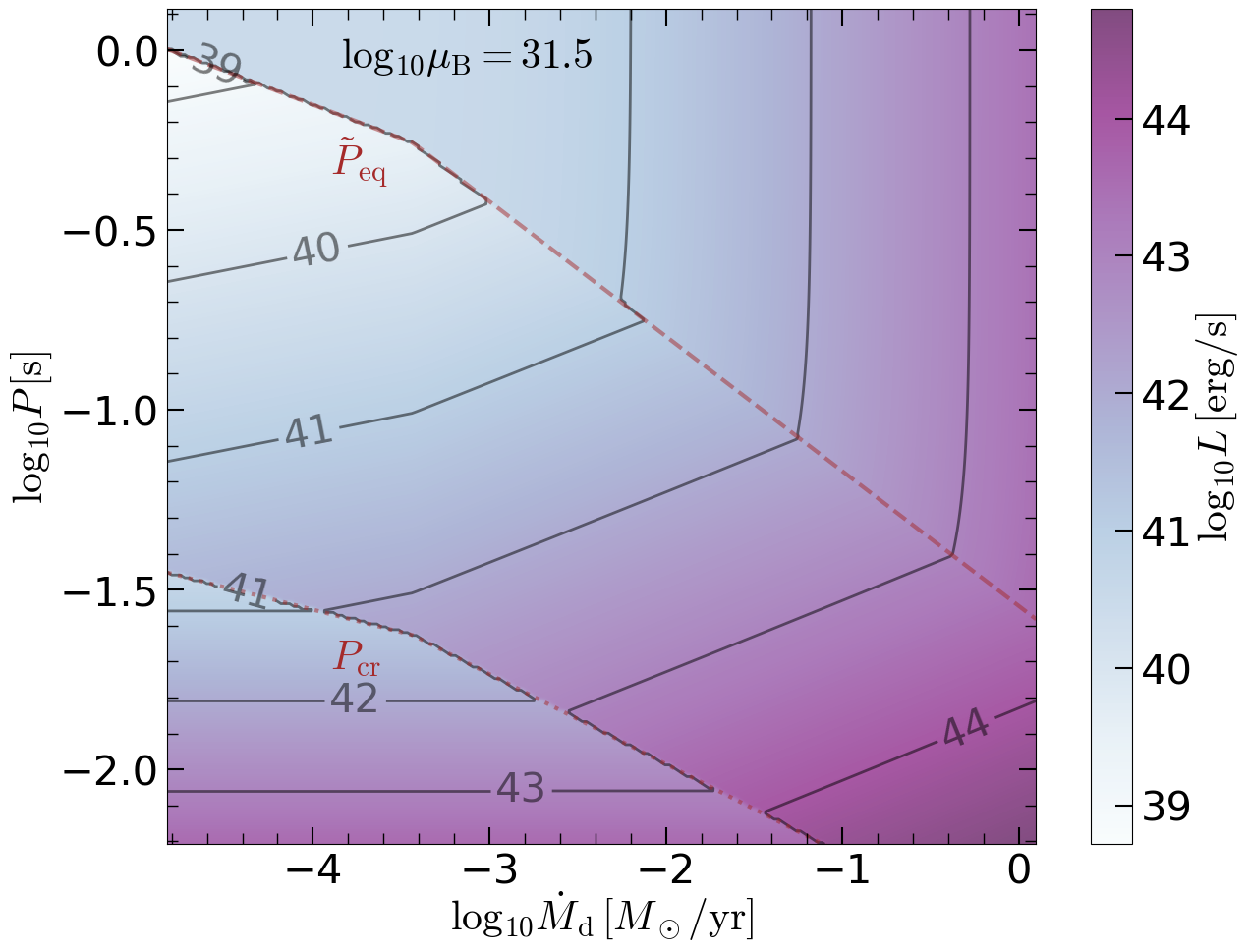}
\caption{Total power from the neutron star-disk system as a function of the accretion rate in the outer disk $\dot{M}_{\rm d}$ and the spin period $P=2\pi/\Omega$. We fix the following parameters: magnetic dipole moment $\muB=10^{31.5}\rm\, G\,cm^3$ (similar to the Crab Pulsar), index for the radial scaling of the accretion rate $p=0.5$, and the outer disk radius $\rd=R_\odot$. The system may be in three different states: pulsar state ($P<P_{\rm cr}$ region), propeller state ($P_{\rm cr} < P< \tilde{P}_{\rm eq}$ region), and accretor state ($P>\tilde{P}_{\rm eq}$ region).
}
\label{fig:total_acc_power}
\end{figure}

The total power of the system as a function of the accretion rate of the outer disk $\dot{M}_{\rm d}$ and the neutron star's spin period is shown in Fig. \ref{fig:total_acc_power}. We find that, for a modest magnetic dipole moment $\muB = 10^{31.5}\rm\, G\,cm^{-3}$ (or $B_{\rm ns}\simeq 3\times 10^{13}\rm\, G$), spin period $P\sim 30\rm\, ms$ (similar to the Crab Pulsar), and accretion rates $\dot{M}_{\rm d}\gtrsim 10^{-2}\rm\,M_\odot\,yr^{-1}$, it is possible to obtain very high powers $L\gtrsim 10^{42.5}\rm\, erg\,s^{-1}$ in the propeller state. Such accretion rates are achievable based on the interactions between the neutron star and the inflated envelope of the companion star (see eq. \ref{eq:acc_rate_outer_disk}). In the propeller state, most of the power comes from the rotational energy of the neutron star (meaning that $L_{\rm w}$ dominates over $L_{\rm m}$ and $L_{\rm d}$), and the accretion flow taps the rotational energy faster than the magnetic dipole emission. 

\begin{figure}
\centering
\includegraphics[width=0.48\textwidth]{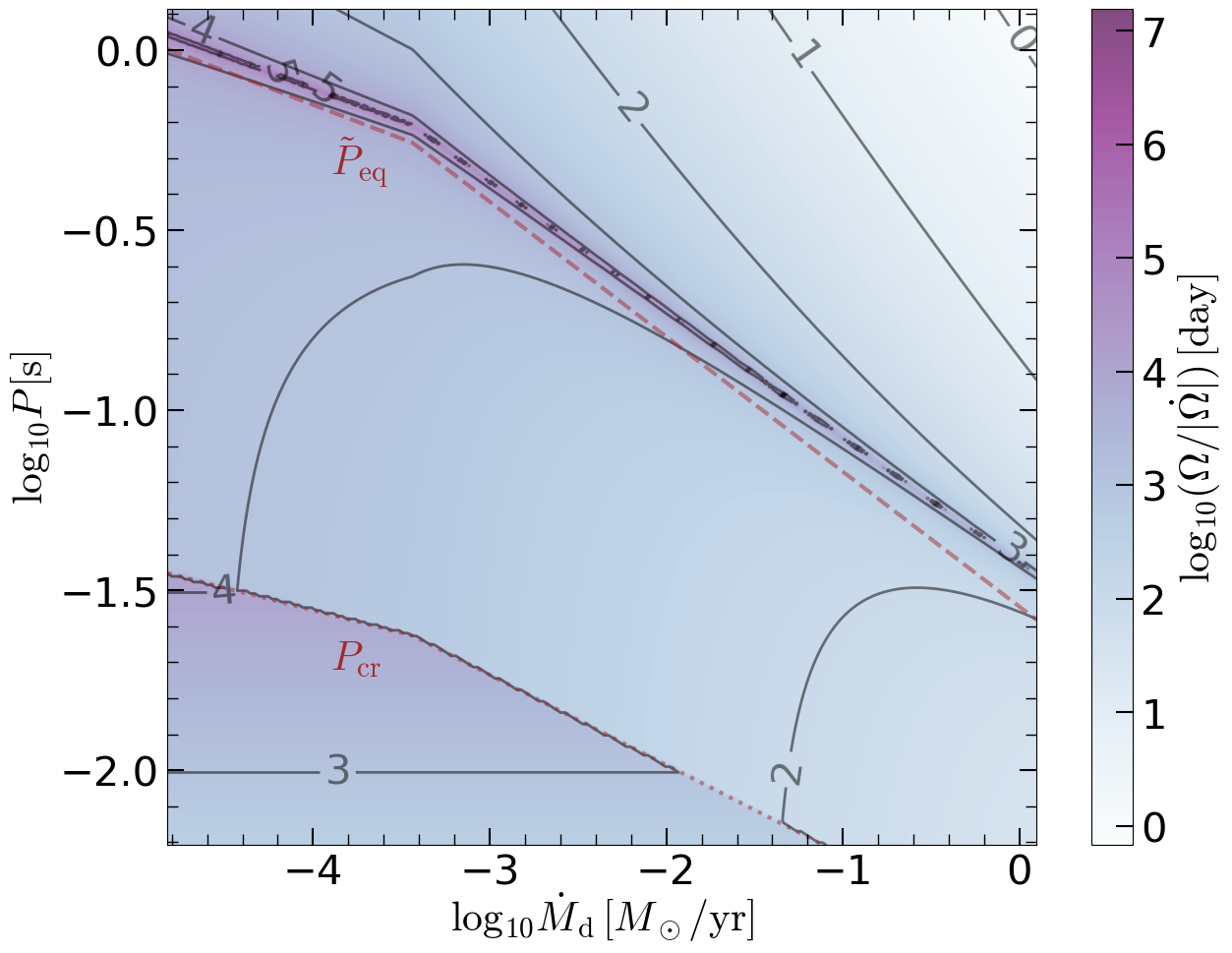}
\caption{The timescale for spin evolution $\Omega/|\dot{\Omega}|$ for the same system shown in Fig. \ref{fig:total_acc_power}. Note that, due to a non-zero electromagnetic spin-down torque $-L_{\rm m}/\Omega$ (eq. \ref{eq:spin_torques}), the spin equilibrium at $\dot{\Omega}=0$ does not coincident with the dashed line for $P=\tilde{P}_{\rm eq}$.
}
\label{fig:spindown_timescale}
\end{figure}

The timescale for the neutron star's spin evolution $\Omega/|\dot{\Omega}|$ is shown in Fig. \ref{fig:spindown_timescale}. For $\mu_{\rm B} = 10^{31.5}\rm\, G\,cm^3$, the spin evolution timescale is generally longer than $10^2\rm\, d$ for accretion rates $\dot{M}_{\rm d}\lesssim 0.1 \msunyr$, so spin evolution occurs on timescales comparable or longer than the evolutionary timescale of the accretion rate.



We define the accretion efficiency of the system as
\begin{equation}\label{eq:acc_efficiency}
    \eps \equiv {L_{\rm acc}\over \dot{M}_{\rm d}c^2} \equiv {L_{\rm w}+L_{\rm ns}+L_{\rm d}\over \dot{M}_{\rm d}c^2},
\end{equation}
which does not include the electromagnetic power $L_{\rm m}$ (which is negligible in our calculations).
From Fig. \ref{fig:total_acc_power}, we see that the accretion luminosity $L_{\rm acc}$ is maximized when $P = P_{\rm cr}$ or the light cylinder radius coincide with the Alfv\'en radius $\rlc=\rA$. When this occurs, the gas that reaches down to the Alfv\'en radius is flung away at the corotational speed which is close to the speed of light. To correctly obtain the system's total power, a fully relativistic calculation is required --- our Newtonian treatment is only applicable when the spin period is slightly longer than the critical period $P_{\rm cr}$.
At $P = P_{\rm cr}$, the ratio between the electromagnetic power and the propeller wind power is given by\footnote{At $P<P_{\rm cr}$, the disk truncates at $\rA > \rlc$ and the accreting gas is not forced into corotation with the magnetosphere, so the propeller wind disappears. This might be an artifact of our current model for the propeller state. Future theoretical works should address if this discontinuity is real. }
\begin{equation}\label{eq:Lm_over_Lw_at_Pcr}
\begin{split}
    {L_{\rm m}\over L_{\rm w}}(P=P_{\rm cr}) &= 2\sqrt{2G\Mns /(\rlc c^2)}\\
    &=2\sqrt{2\rg/\rlc} = 0.11 (P/30\mr{ms})^{-1/2},
\end{split}
\end{equation}
where $\rg = G\Mns/c^2$ is the gravitational radius of the neutron star. The ratio between the disk accretion power and the propeller wind power is given by
\begin{equation}
    {L_{\rm d}\over L_{\rm w}}(P=P_{\rm cr}) = \rg/\rlc = 1.4\times10^{-3} (P/30\mr{ms})^{-1}.
\end{equation}
Therefore, for sufficiently slow pulsars $P\gtrsim 10\, \mr{ms}$, we find that the total power is dominated by the propeller wind $L \approx L_{\rm w} \approx \dot{M}(\rlc) c^2$ when $P=P_{\rm cr}$. Thus, the maximum accretion efficiency is given by
\begin{equation}\label{eq:max_acc_efficiency}
    \eps_{\rm max} \approx \lrb{\rA\over \rd}^p = 3.9\%\, \dot{M}_{\rm d,-1}^{-{1\over 8}} \mu_{\rm B,31.5}^{1\over 4} \lrb{\rd\over 0.3R_\odot}^{-{7\over 16}},
\end{equation}
where we have used $\dot{M}_{\rm d} = 10^{-1}\,\msunyr\, \dot{M}_{\rm d,-1}$, $\mu_{\rm B}=10^{31.5}{\rm\, G\,cm^3}\, \mu_{\rm B,31.5}$, and $p=0.5$. For a more optimistic case of $p=0.3$, we obtain $\eps_{\rm max}\approx 12.7\% \dot{M}_{\rm d,-1}^{-0.08} \mu_{\rm B,31.5}^{0.16} (\rd/0.3R_\odot)^{-0.28}$.

The above estimate shows that $\eps_{\rm max}$ is relatively insensitive to the highly uncertain accretion rate $\dot{M}_{\rm d}$ and magnetic dipole moment $\mu_{\rm B}$. Since our Bondi-capture picture gives a robust disk radius of $\rd \sim 0.3 R_\odot$ (see eq. \ref{eq:disk_radius}), we conclude that the accretion system has a maximum efficiency of a few percent (for $p=0.5$), which is achieved when the system is in the propeller state near the boundary of the pulsar state $P\simeq P_{\rm cr}$.
Away from the boundary of $P=P_{\rm cr}$ but still in the propeller state, the efficiency is given by
\begin{equation}\label{eq:propeller_efficiency_general}
    \eps/\eps_{\rm max} = \lrb{\rA/\rlc}^2 = (P/P_{\rm cr})^{-2}.
\end{equation}
We see that the accretion efficiency drops as $\eps \propto P^{-2}$ for longer spin periods if we keep $\dot{M}$, $\rd$, and $\mu_{\rm B}$ fixed. On the other hand, if we keep $P$, $\rd$, and $\muB$ fixed, the accretion efficiency scales as $\eps\propto \dot{M}_{\rm d}^{-1/2}$ (for $p=0.5$).

On the other hand, in the limit of an extremely slow spin or weak magnetic dipole moment, we obtain the minimum accretion efficiency for $p=0.5$,
\begin{equation}
    \eps_{\rm min} = \lrb{R_{\rm ns}\over \rd}^p {r_{\rm g}\over R_{\rm ns}} \simeq 1.3\times10^{-3} \lrb{\rd/0.3R_\odot}^{-1/2}.
\end{equation}
We note that the minimum efficiency is likely too small to power the second peak in SN2022jli, unless the accretion power-law index is smaller $p<0.5$ or the captured mass per orbit is as large as $\Delta M_{\rm cap}\sim 10^{-3}M_\odot$.




In the next section, we discuss how the accretion power may be converted into observable radiation, while keeping our discussion agnostic to the detailed disk outflow launching mechanisms (propeller- or viscously-driven wind).




\section{Emission from shocks}\label{sec:shock_emission}

\begin{figure}
\centering
\includegraphics[width=0.42\textwidth]{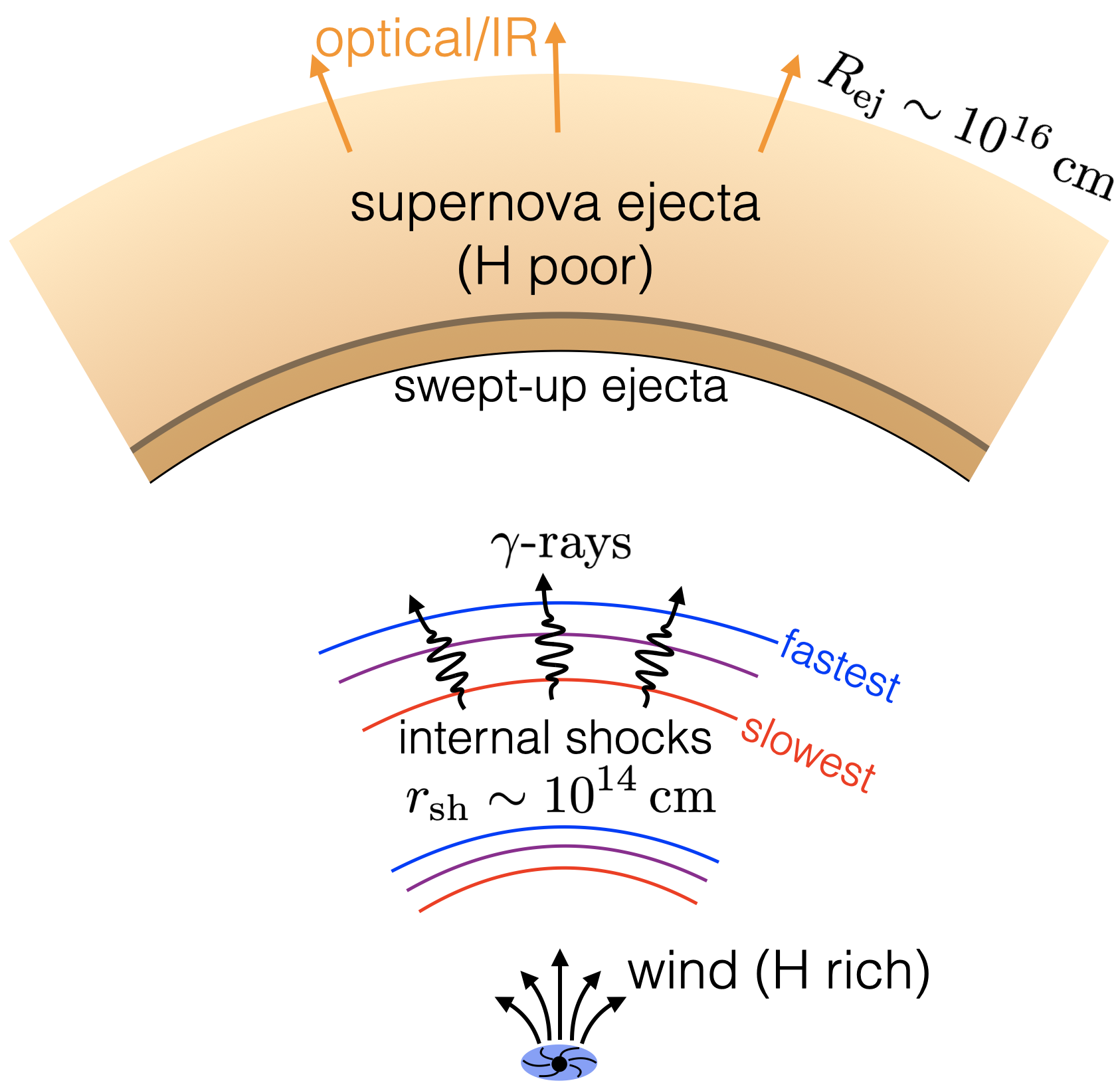}
\caption{Geometry of the system on lengthscales from $10^{14}$ to $10^{16}\rm\,cm$. 
}
\label{fig:shock_engine}
\end{figure}

The accretion-driven outflow launched near disk radius $r$ has mass loss rate $\dot{M}(r) = \dot{M}_{\rm d} (r/\rd)^{p}$ and speed $v(r)$.
At large radii $r\gg \rd$, the outflow will be radially stratified due to the velocity difference. The fastest outflow that carries the majority of the accretion power will collide with slower material on its way outward. There are two locations for energy dissipation: (1) for a bound post-supernova orbit, disk accretion occurs periodically on the orbital timescale, so the slower outflow from an earlier episode will lie ahead of the freshly launched faster outflow; (2) the cumulative disk outflow will collide with the supernova ejecta at much larger radii. The geometry of the system is schematically shown in Fig. \ref{fig:shock_engine}.
These collisions create shocks that dissipate the outflow's kinetic energy into heat. If the cooling timescale of the shock-heated gas is shorter than the dynamical expansion timescale, such shocks have a high radiative efficiency, so bright emission will be produced. In the following, we first focus on the internal shocks that form due to the collisions between adjacent episodes of disk outflows and then briefly comment on the radiative efficiency of the wind-ejecta shocks. We consider the cooling of the thermal and non-thermal electrons in \S \ref{sec:thermal_electrons} and \ref{sec:non_thermal_electrons}, respectively.

\subsection{Cooling of thermal electrons}\label{sec:thermal_electrons}

The fastest outflow, originated from near the Alfv\'en radius, can reach up to a significant fraction of $c$ in the extreme limit (if $\rA \simeq \rlc$). The slowest outflow is launched from the outer disk near $r\sim \rd \sim 0.3 R_{\odot}$, so the minimum speed is of the order $v_{\rm min} \sim \sqrt{G\Mns/\rd} \sim 10^8 \rm\, cm\,s^{-1}$ (roughly the local Keplerian speed). For a post-supernova orbital period of $P = 10 P_{\rm 10d}\rm\, day$, the internal shocks occur at a characteristic radius
\begin{equation}
    r_{\rm sh} \simeq v_{\rm min} P \sim 10^{14}\mr{\,cm}\, P_{\rm 10d}.
\end{equation}

The total gas mass ejected in each binary orbit is of the order $\Delta M_{\rm cap}$, and the total energy carried by the fastest outflow is of the order
\begin{equation}
    \Delta E \sim \eps\, \Delta M_{\rm cap} c^2 = 1.8\times10^{48}\mr{\,erg}\, \eps_{-2} {\Delta M_{\rm cap}\over 10^{-4}M_\odot}, 
\end{equation}
where $\eps$ is the accretion efficiency, and we have taken $\eps=10^{-2}\eps_{-2}$ (eq. \ref{eq:acc_efficiency}). We assume that an order-unity fraction of the ejected mass is swept up by the forward shock and expands at speed 
\begin{equation}
    v_{\rm exp} \simeq 0.1 c\, \eps_{-2}^{1/2}.
\end{equation}
Thus, the dynamical expansion timescale of the shock-heated gas is given by
\begin{equation}\label{eq:dynamical_time}
\begin{split}
    t_{\rm dy} &= \max\lrsb{t_{\rm vis}, {r_{\rm sh}/v_{\rm exp}}} \\
    &= \max\lrsb{0.3\mr{\,d}\, {(\rd/0.3R_\odot)^{3/2}\over \alpha h^{2}/0.01}, 0.4\mr{\,d}\, r_{\rm sh,14} \eps_{-2}^{-1/2}},
\end{split}
\end{equation}
where we have assumed that most of the accretion energy is injected on the viscous timescale $t_{\rm vis}$ (eq. \ref{eq:viscous_time}).

Let us denote the shock speed as $\beta c$, which is measured in the frame of the unshocked gas. Below, we take $\beta = 0.1\beta_{0.1}$ to be a typical fiducial value for the forward shock that propagates into the slower, denser outflow component. We note that the reverse shock crossing the faster outflow may have a higher $\beta$, and the cooling of reverse shock-heated gas (dominated by inverse-Compton scattering) requires a strong radiation field that is provided by free-free emission by the much denser forward shock-heated gas. The following discussion on inverse-Compton cooling applies to electrons heated by both forward and reverse shocks.

For a given shock speed $\beta c$, the shock-heated gas has temperature
\begin{equation}
    T_{\rm sh} = {3\mu \mproton (\beta c)^2\over 16 \kB} = 1.2\times10^{10}\mr{\,K}\, \beta_{0.1}^2.
\end{equation}
We assume that electrons and ions reach an equilibrium\footnote{If electron temperature $T_{\rm e}$ is less than the ion temperature $T_{\rm i}$ \citep[numerical simulations suggest $0.2\lesssim T_{\rm e}/T_{\rm i}\lesssim 1$, e.g.,][]{Spitkovsky:2008aa, Kato:2010aa, Park:2015aa, Crumley:2019aa}, we should roughly replace $T_{\rm sh}$ by $T_{\rm e}$. The thermal electrons still cool efficiently, but the overall radiative efficiency of internal shocks will be smaller by a factor of a few.} with the same kinetic temperature $T_{\rm sh}$. The Thomson scattering optical depth for the slowest disk outflow heated by the forward shock is given by
\begin{equation}\label{eq:scattering_optical_depth_internal_shocks}
    \tau_{\rm s} \simeq {\kappa_{\rm T} \Delta M_{\rm cap} \over 4\pi r_{\rm sh}^2} \simeq 0.5 {\Delta M_{\rm cap}\over 10^{-4}M_\odot} r_{\rm sh,14}^{-2},
\end{equation}
where we have taken the Thomson opacity $\kappa_{\rm T}\approx 0.34\rm\,cm^2\,g^{-1}$ for solar metallicity. Klein-Nishina effects further suppress the scattering cross-section, making the shock-heated region even more optically thin.

The relevant processes for electron cooling are free-free emission and inverse-Compton scattering. For sufficiently high shock speeds $\beta\gtrsim 0.1$ and low scattering optical depths $\tau_{\rm s}\lesssim 1$, inverse-Compton cooling generally dominates over free-free cooling \citep[see][who treat the Comptonization of free-free photons in detail]{Margalit:2022aa}.

The hydrogen number density in the forward shock-heated region is roughly given by
\begin{equation}\label{eq:proton_density_shocked_region}
    n \sim {\Delta M_{\rm cap}\over V_{\rm sh} \mproton} \sim 1\times 10^{11}\mr{\,cm^{-3}}\, r_{\rm sh,14}^{-3} {\Delta M_{\rm cap}\over 10^{-4}M_\odot},
\end{equation}
where we have taken the volume of the shock-compressed gas to be $V_{\rm sh}\sim r_{\rm sh}^3$. The total free-free luminosity is then given by
\begin{equation}
\begin{split}
    L_{\rm ff} &= 4\pi j_{\rm ff} V_{\rm sh} f_{\rm rel,ff}\\
    &\simeq 3\times10^{42}\mr{\,erg\,s^{-1}}\, \beta_{0.1} r_{\rm sh,14}^{-3} {f_{\rm rel}\over 2} \lrb{\Delta M_{\rm cap}\over 10^{-4}M_\odot}^2,
\end{split}
\end{equation}
where $f_{\rm rel}$ is a correction factor that smoothly connects the non-relativistic to the relativistic regimes
\begin{equation}
    f_{\rm rel, ff} \simeq 1 + {1.3 (T_{\rm sh}/10^{10}\rm\, K)\over 1 + 0.6 (T_{\rm sh}/10^{10}\rm\, K)^{1/2}}.
\end{equation}
Then, the photon energy density photons from free-free emission is given by (in the optically thin limit)
\begin{equation}
\begin{split}
    U_{\rm ff} &\simeq {L_{\rm ff}/(4\pi r_{\rm sh}^2 c)}\\
    &\simeq 800\mr{\,erg\,cm^{-3}} \beta_{0.1} r_{\rm sh,14}^{-5} {f_{\rm rel}\over 2} \lrb{\Delta M_{\rm cap}\over 10^{-4}M_\odot}^2.
\end{split}
\end{equation}
Although free-free emission alone is unable to efficiently cool the shock-heated electrons, inverse-Compton scattering of these seed photons is more efficient and the corresponding cooling time is 
\begin{equation}\label{eq:IC_cooling_timescale}
\begin{split}
    t_{\rm IC} &
    = {3 \me c^2 \over 8 U_{\rm ff} \sigma_{\rm T} c} f_{\rm rel,IC}^{-1}\\
    &\simeq 0.02\mr{\,d}\, \lrb{f_{\rm rel, IC}\over 10}^{-1} \lrb{U_{\rm ff}\over 10^3\rm\,erg\,cm^{-3}}^{-1},
\end{split}
\end{equation}
where $f_{\rm rel,IC}$ is a relativistic correction factor,
\begin{equation}
    f_{\rm rel,IC} \simeq 1 + {4\kB T_{\rm sh} \over \me c^2} = 1 + 8.1\beta_{0.1}^2.
\end{equation}
The above estimate does not take into account the effects of electron recoil or Klein-Nishina suppression of the scattering cross-section. We expect the scattering of the photons near the maximum frequency of the free-free spectrum $h\nu_{\rm max} \sim \kB T_{\rm sh}\sim 1\rm\, MeV$ to be Klein-Nishina suppressed, which will reduce the inverse-Compton power by a factor of a few to 10. However, since the Compton-$y$ parameter of the forward shock-heated region is much greater than unity,
\begin{equation}
    y \simeq 16\lrb{\kB T_{\rm sh}\over \me c^2}^2 \tau_{\rm s} \simeq 35\, \beta_{0.1}^4 r_{\rm sh,14}^{-2} {\Delta M_{\rm cap}\over 10^{-4}M_\odot},
\end{equation}
the lower energy free-free photons will be significantly upscattered. As long as the energy density of soft photons at $h\nu\ll \me c^2$ exceeds $\sim\!50\rm\, erg\,cm^{-3}$, inverse-Compton scattering will efficiently cool the thermal electrons heated by both reverse shock and forward shock on a timescale $t_{\rm IC}$ that is shorter than the dynamical expansion time $t_{\rm dy}$ (eq. \ref{eq:dynamical_time}).

Synchrotron emission may also provide significant significant source photons for inverse-Compton cooling.
Although the synchrotron photons by thermal electrons are mainly in the radio band and thus are self-absorbed, higher-frequency synchrotron emission by non-thermal electrons will be inverse-Compton scattered by thermal electrons. However, we do not discuss the synchrotron self-Compton cooling in detail here.

Based on the above discussion, we conclude that the internal shocks between adjacent episodes of mass ejections from the accretion disk are radiatively efficient --- an order unity fraction of the kinetic energy of the fastest outflow will be converted into radiation. Most of the emitted energy is carried by photons near $h\nu\sim \kB T_{\rm sh}\sim \rm\, 1\,MeV$ (limited by electron recoiling). As long as the MeV photons efficiently deposit their energy into the supernova ejecta via Compton heating \citep[which is the case in the first 100--300 days for typical Type Ib/c supernovae, e.g.,][]{Wheeler:2015aa}, we expect bright optical emission to be produced.

The discussion above focused on the emission from internal shocks. In the following, we briefly discuss another location for energy dissipation --- the shock driven into the supernova ejecta, whose mass is mostly located at much larger radii $\gg r_{\rm sh}$.

Let us denote the remaining energy of the accretion-driven outflow as $E_{\rm w}$, which is comparable to the cumulative energy output from accretion, as the radiative losses of internal shocks can only deplete the thermal energy of the shock-heated gas but not the kinetic energy of the bulk motion. We assume a uniform unperturbed ejecta density profile, with the location of the shock-heated ejecta at 
\begin{equation}
\begin{split}
    r_{\rm sh,ej}(t) &\simeq \lrb{E_{\rm w}\over E_{\rm ej}}^{1/5} R_{\rm ej}(t)\\
    &\simeq 3\times10^{15}\mr{\,cm}\,   {t\over 100\mr{\,d}} {v_{\rm ej}\over 10^9\mr{\,cm\,s^{-1}}} \lrb{E_{\rm w,49}\over E_{\rm ej,51}}^{1/5},
\end{split}
\end{equation}
where $R_{\rm ej} = v_{\rm ej} t$ is the characteristic radius of the ejecta, $v_{\rm ej} \simeq \sqrt{2E_{\rm ej}/M_{\rm ej}}$, and $t$ is the time since explosion. Using the same method as above, we estimate the free-free cooling timescale of the shock-heated ejecta to be 
\begin{equation}
    t_{\rm ff,ej} \simeq \mr{10\,d}\, (t/100\mr{\,d})^3 {8\over \bar{Z}} { E_{\rm ej,51}^{3/2} \over \lrb{M_{\rm ej}/3M_\odot}^{5/2}},
\end{equation}
where $\bar{Z}\simeq 8$ is the mean charge number of the ions for an oxygen-dominated composition that is fully ionized.
We see that the shock propagating into the supernova ejecta is a radiative one due to rapid cooling ($t_{\rm ff,ej} \ll t$). The typical shock velocity is given by $(r_{\rm sh,ej}/R_{\rm ej})v_{\rm ej}\sim 0.01c$, so the free-free emission will be in the X-ray band at $h\nu\lesssim 10\rm\, keV$ --- these photons will deposit a fraction of their energies into the supernova ejecta by bound-free absorption (mainly by Fe group elements).

We conclude that both the internal shocks within the episodic accretion outflows (occurring at radii $r_{\rm sh}\sim 10^{14}\rm\,cm$) and the shock propagate into the supernova ejecta (at radii $r_{\rm sh,ej}\sim \mr{\,few}\times10^{15}\rm\,cm$) can efficiently dissipate and then radiate away the kinetic energy of the accretion outflow. The radiation will then be reprocessed into the optical band by the supernova ejecta. Note that only the emission from internal shocks will be modulated by the orbital period of the post-supernova binary. In the optical band, the amplitude of periodic modulation is diluted as the light-crossing time of the ejecta, 
\begin{equation}
    t_{\rm lc} = R_{\rm ej}/c \simeq 3\mr{\,d}\, (t/100\mr{\,d}) (v_{\rm ej}/10^9\mr{\,cm\,s^{-1}}),
\end{equation}
and is much longer than the timescale over which the emission from internal shocks are produced ($t_{\rm dy}\lesssim 1\rm\,day$, eq. \ref{eq:dynamical_time}). At sufficiently late times such that $t_{\rm lc} \gtrsim P$, the reprocessed emission will be severely smeared and the periodic modulation in the optical band disappears.

\subsection{High-energy emission by non-thermal particles}\label{sec:non_thermal_electrons}

The internal shocks between adjacent episodes of disk outflows are collisionless because the gas has low scattering optical depth (eq. \ref{eq:scattering_optical_depth_internal_shocks}). We expect such collisionless shocks to accelerate non-thermal electrons and protons, which take fractions $\eps_{e}$ and $\eps_{p}$ of the thermal energy in the shock-heated region, respectively. Previous theoretical and observational work on shock-accelerated cosmic rays suggest $\eps_{p}\sim 0.1$, but  $\eps_{e}$ remains uncertain and might be much lower \citep[likely due to inefficient injection for diffusive shock acceleration, e.g.,][]{Park:2015aa}.

In the following, we consider the radiation from the non-thermal particles accelerated by internal shocks.



Without considering radiative losses, the maximum energy for a cosmic ray particle of charge $Ze$ is given by the \citet{Hillas:1984aa} criterion (the Larmor radius $\sim r_{\rm sh}$),
\begin{equation}
    \mc{E}_{\rm max} \sim {eB r_{\rm sh}} \sim 1\mr{\,EeV} \,Z\, \lrb{{\eps_{\rm B,-4}\eps_{-2} \over r_{\rm sh,14}} {\Delta M_{\rm cap}\over 10^{-4}M_\odot}}^{1/2},
\end{equation}
where we have taken the energy density of the magnetic field $B^2/8\pi$ in the shocked gas to be a fraction $\eps_{\rm B}=10^{-4}\eps_{\rm B,-4}$ of the total energy density $U \sim \eps \Delta M_{\rm cap} c^2/V_{\rm sh}$. Another criterion for the maximum energy of electrons is set by synchrotron losses. Setting the gyration of the angular frequency to be equal to the inverse of the synchrotron cooling time, we obtain the following
\begin{equation}
\begin{split}
    \mc{E}_{\rm e,max} &\sim \sqrt{e\over 6\pi B\sigma_{\rm T}} {\me c^2} \\
    &\sim 0.4\mr{\,TeV}
    \lrb{
    {\eps_{\rm B,-4}\eps_{-2} \over r_{\rm sh,14}^3} {\Delta M_{\rm cap}\over 10^{-4}M_\odot}
    }^{1/2}.
\end{split}
\end{equation}
We see that cosmic ray protons can reach up to $\sim\!\rm EeV$ whereas the energies of cosmic ray electrons are limited to be sub-TeV.

The cooling of ultra-relativistic electrons is dominated by synchrotron emission, as Compton scattering is suppressed by Klein-Nishina effects. This means that the emission from cosmic ray electrons will be a power-law extending up to the synchrotron burn-off limit at $9\me c^2/(4\alpha_{\rm f})=160\rm\,MeV$ \citep{dejager96}, where $\alpha_{\rm f}$ is the fine-structure constant. 


The cooling of cosmic ray protons is dominated by inelastic collisions with other non-relativistic protons (hereafter $pp$ collisions). The cross-section for $pp$ collisions above an incident kinetic energy of a few GeV is about $\sigma_{pp}\gtrsim 3\times10^{-26}\rm\, cm^{-2}$ \citep{Kamae:2006aa}, with a logarithmic increase at higher energies. The mean-free time between adjacent $pp$ collisions is given by
\begin{equation}
    t_{pp} = {1\over n\sigma_{pp}c} \lesssim 0.1\mr{\,d}\, n_{11}^{-1}.
\end{equation}
The fact that $t_{pp}$ is much shorter than the dynamical expansion time $t_{\rm dy}$ (eq. \ref{eq:dynamical_time}) means that protons with $\mc{E}_{\rm kin}\gtrsim \mr{GeV}$ will efficiently convert their kinetic energies into pions, $\pi^0$ and $\pi^\pm$. In about $1/3$ of the cases, neutral pions $\pi^0$ will be produced and they subsequently decay into photons in a broad energy band from 100 MeV up to about 10 PeV, with the upper limit due to maximum proton energy limited by the short timescale for $pp$ collisions ($\mc{E}_{p,\rm max}\sim 100\rm\, PeV$ for our fiducial parameters). Most of the radiation energy will be in the GeV band. In about 2/3 of the cases, charged pions $\pi^{\pm}$ will be produced and they subsequently decay into $e^\pm$ and neutrinos (with roughly equal energy between the two decay products). The $e^\pm$ will then cool by synchrotron emission and produce low-energy photons.
Including the photon emission from $\pi^0$ decays and cooling of secondary $e^\pm$, we find that the radiative efficiency of cosmic ray protons to be rather high, $\eps_{p\rightarrow\gamma} \sim 2/3$.

For cosmic ray proton energy partition $\eps_p=0.1\eps_{p,-1}$ and the efficiency of GeV gamma-ray emission $\eps_{\rm GeV}\sim 1/3$, we estimate the orbit-averaged luminosity in the GeV band to be
\begin{equation}\label{eq:GeV_luminosity}
\begin{split}
    L_{\rm GeV} &= {\eps_{\rm GeV}\,\eps_p \eps\, \Delta M_{\rm cap} c^2\over P}\\
    &= 7\times10^{40}\mr{\,erg\,s^{-1}} {\eps_{\rm GeV}\over 1/3} \eps_{p,-1} \eps_{-2} {\Delta M_{\rm cap}\over 10^{-4}\Msun} {10\mr{\,d}\over P}.
\end{split}
\end{equation}
The secondary neutrinos, which take away a fraction $\eps_{p\rightarrow\nu} \sim 1/3$ of the cosmic ray protons' energy, will also have a broad energy spectrum between 100 MeV and 10 PeV. These neutrinos, with a luminosity that is comparable to $L_{\rm GeV}$ above, are a smoking-gun signature of $pp$ collisions in our model. Future observations of the PeV neutrino emission \citep[by Icecube,][]{aartsen17_icecube} will provide a decisive test of our model.

\begin{figure}
\centering
\includegraphics[width=0.48\textwidth]{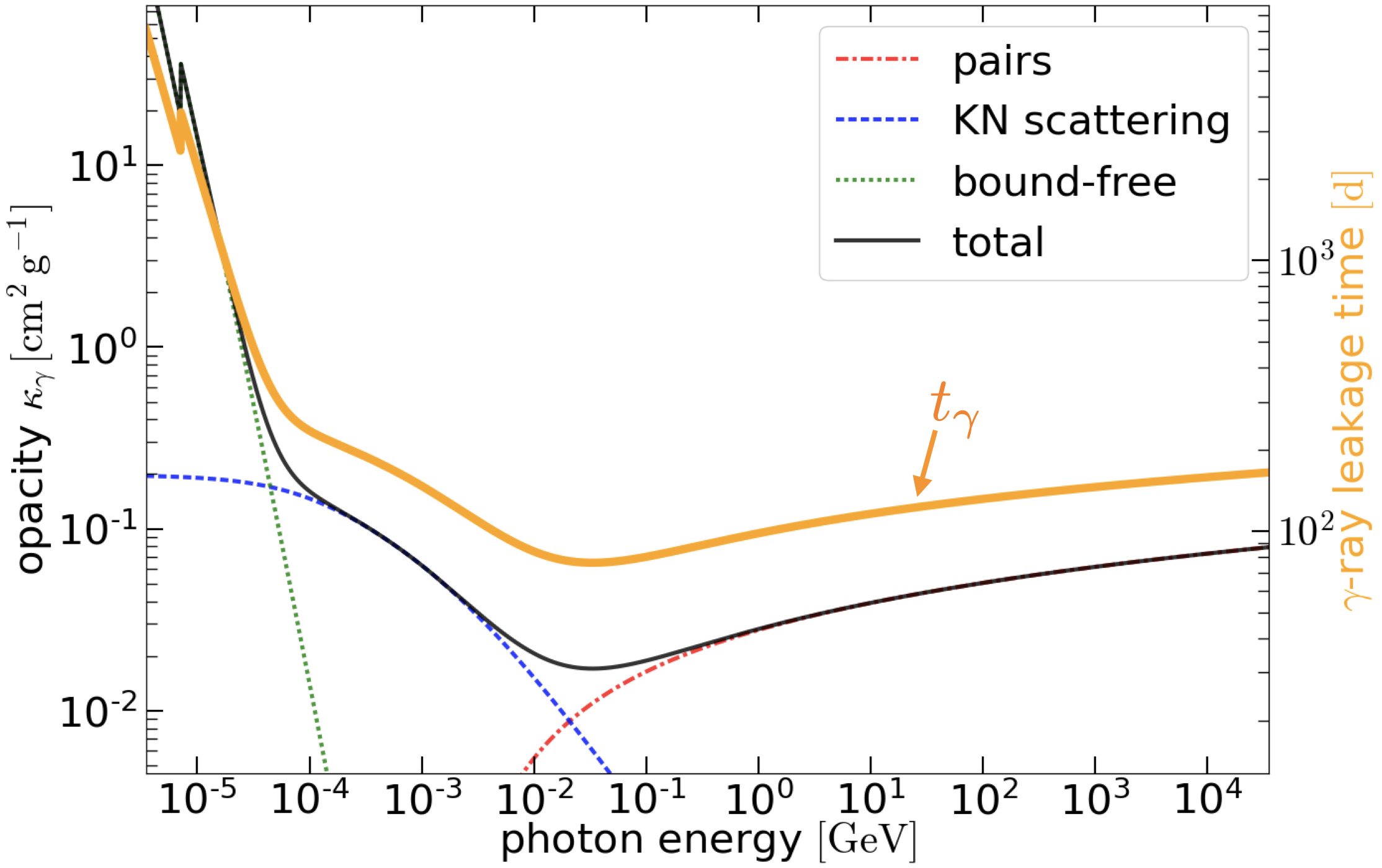}
\caption{Opacity at high photon energies for a mock supernova ejecta composition: 10\% He, 30\% C, 40\% O, 10\% Ne, 5\% Si, and 5\% Fe-group. The sources of opacity include Bethe-Heitler pair production \citep{Hubbell:1980aa} in red dash-dotted line, Klein-Nishina (KN) electron scattering \citep{Rybicki:1986aa} in blue dashed line, and bound-free (bf) absorption \citep{Verner:1996aa} in green dotted line, and the black solid line shows the sum of all. The thick orange line shows the $\gamma$-ray leakage time $t_{\gamma}$ in days (eq. \ref{eq:gamma_leakage_time}) on the right axis, for ejecta mass $M_{\rm ej}=3M_\odot$ and energy $E_{\rm ej}=10^{51}\rm\, erg$. Here, we assume a spatially uniform composition, whereas realistic ejecta has more centrally concentrated Si and Fe distributions, so $t_{\gamma}$ should be slightly longer than our estimate.
}
\label{fig:gamma_leakage_time}
\end{figure}

The escaping of the high-energy photons depends on the opacity of the supernova ejecta. In Fig. \ref{fig:gamma_leakage_time}, we show the opacity as a function of photon energy for a mock supernova ejecta composition motivated by the simulations by \citet[][their 5p11 model]{Dessart:2015ab}. The sources of opacities include bound-free transitions, Klein-Nishina scattering, and pair production due to collisions with nuclei and electrons. For the bound-free opacity, we assume all species are in their atomic form, but the result at $\mc{E}_\gamma \gtrsim \rm keV$ is not sensitive to the ionization state of outer-shell electrons. Additionally, the opacity at $\mc{E}_\gamma\gtrsim 0.1\rm \, MeV$ is not sensitive to the detailed mass composition.

For a homologously expanding ejecta with a uniform density profile, the time at which the optical depth for centrally produced $\gamma$-rays reaches $\tau_\gamma=1$ is given by
\begin{equation}\label{eq:gamma_leakage_time}
    t_\gamma = \lrb{{9\over 40\pi} {\kappa_\gamma M_{\rm ej}^2\over E_{\rm ej}}}^{1/2} \simeq 100 \mr{\,d}\, \kappa_{\gamma,0.03}^{1/2} E_{\rm ej,51}^{-1/2} {M_{\rm ej}\over 3M_\odot},
\end{equation}
where $\kappa_\gamma = 0.03 \mr{\,cm^2\,g^{-1}}\,\kappa_{\gamma,0.03}$ is the opacity near photon energy of 1 GeV (Fig. \ref{fig:gamma_leakage_time}). For a more highly centrally concentrated density profile and compositional distribution of heavier elements like Si and Fe, the $\gamma$-ray leakage time may be larger by a factor of $\sim\!2$. Nevertheless, we expect the GeV photons to leak out of the ejecta after a delay of the order 100 days, which is comparable to the timescale $t_{\rm max}$ (eq. \ref{eq:tmax_companion_star}) for the impacted companion star to expand to its maximum radius. This means that it is possible to detect GeV emission from the system by e.g., Fermi LAT \citep{Atwood:2009aa}.

At even higher energies $\mc{E}_\gamma\sim\rm TeV$, it takes longer for the $\gamma$-rays to leak out ($\kappa_\gamma\simeq 0.06 \mr{\,cm^2\,g^{-1}}$ at $1\rm\, TeV$). However, pair production by $\gamma+\gamma\rightarrow e^\pm$ (not included in Fig. \ref{fig:gamma_leakage_time}) becomes important due to the abundant optical photons, so we do not expect to detect TeV photons from the system.
We also do not expect to detect X-ray photons with energies $\lesssim 30 \rm\, keV$, as they are fully trapped
by the ejecta for many years, which is longer than the duration of the accretion from the companion star. 

\subsection{Lightcurve model}

\begin{figure}
\centering
\includegraphics[width=0.48\textwidth]{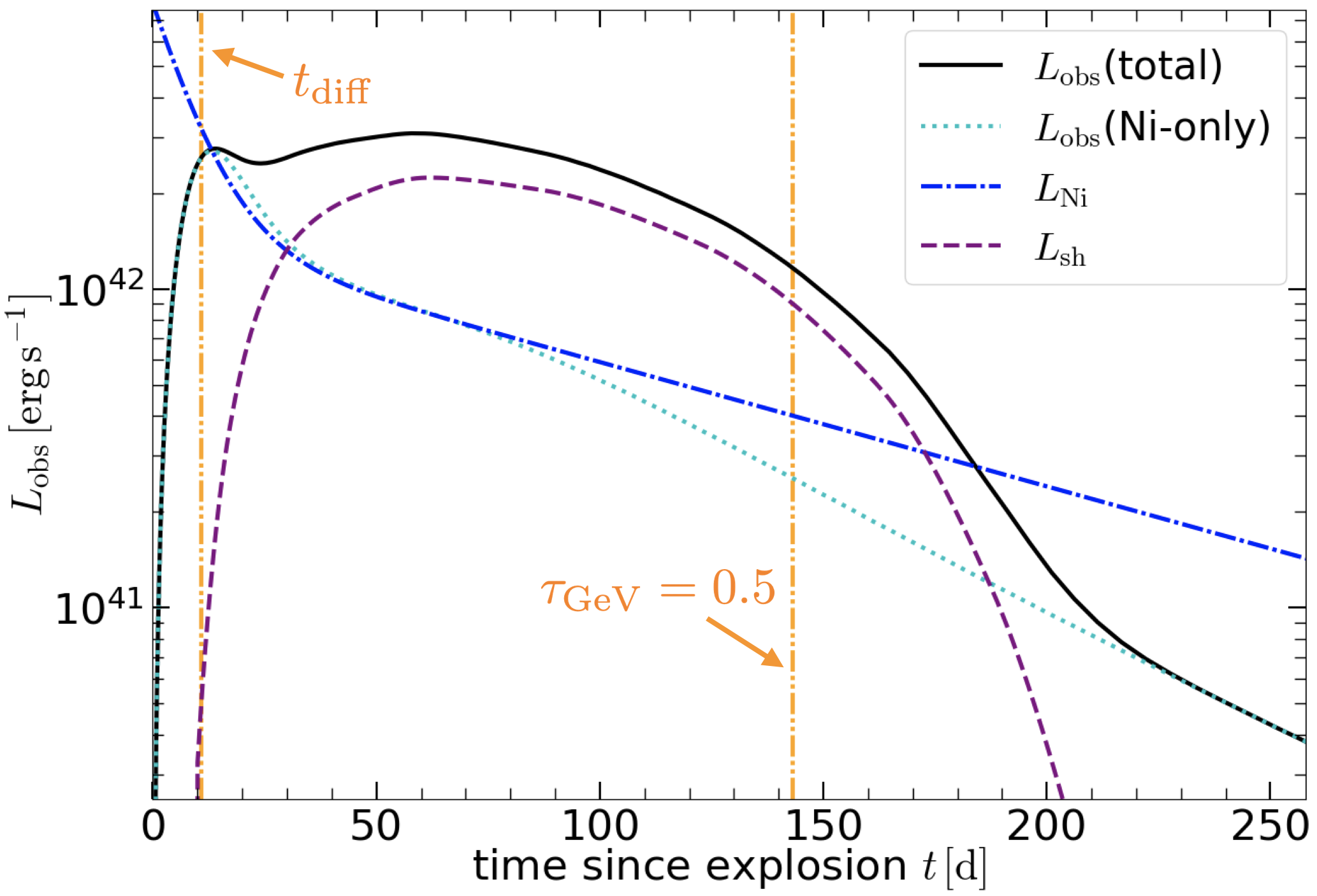}
\caption{Crude model lightcurve with parameters $\eps=3\%$ (accretion efficiency), $P_{\rm orb}=10\rm\,d$, $M_*=2M_\odot$, $a_0=18R_\odot$, $\rp=12R_\odot$, $M_{\rm ej}=3M_\odot$, $M_{\rm Ni}=0.1M_\odot$, $E_{\rm ej}=10^{51}\rm\, erg$ \citep[motivated by][]{Woosley:2021aa}. We mark the time when the optical depth for GeV photons drops to $\tau_{\rm GeV}=0.5$. 
}
\label{fig:lightcurve}
\end{figure}

In this subsection, we attempt to construct a very crude model for the optical lightcurve of the system, leaving to future works the detailed consideration of the periodic ejecta heating due to episodic internal shocks. Our results, as shown in Fig. \ref{fig:lightcurve}, are obtained as follows.

The UV-optical-infrared ``bolometric'' luminosity from the system is given by \citep[][their eq. 36 but with a general heating rate]{Arnett:1982aa}
\begin{equation}
    L_{\rm obs}(t) = 2 \mr{e}^{-x^2} \int_0^x L_{\rm h}(x't_{\rm diff}) \mr{e}^{x'^2} x' \d x',
\end{equation}
where $t$ is the observer's time since the explosion, $L_{\rm h}(t' =x' t_{\rm diff})$ is a centrally located heating rate at time $t'\in (0, t)$, and $x \equiv t/t_{\rm diff}$ is the dimensionless time variable scaled to the diffusion time\footnote{The factor of $0.3$ is for the case of a uniform ejecta and it only depends weakly on the density profile \citep{Arnett:1982aa}.} for optical photons
\begin{equation}
\begin{split}
    t_{\rm diff} &\simeq 0.3\, \lrb{\kappa_{\rm opt} M_{\rm ej}^{3/2} \over E_{\rm ej}^{1/2} c}^{1/2}\\
    &\simeq 11\mr{\,d} \lrb{\kappa_{\rm opt}\over 0.02\mr{\,cm^2/g}}^{1/2} \lrb{M_{\rm ej}\over 3M_\odot}^{3/4} E_{\rm ej,51}^{-1/4},
\end{split}
\end{equation}
where the fiducial value for the opacity for optical photons $\kappa_{\rm opt}$ is motivated by \citet{Wheeler:2015aa, Dessart:2015ab}.

The heating rate includes two components
\begin{equation}
    L_{\rm h} = (1-\mr{e}^{-\tau_{\rm MeV}}) L_{\rm Ni} + L_{\rm sh}.
\end{equation}
The first term comes from the $^{56}\mr{Ni}$ decay chain \citep{Nadyozhin:1994aa}
\begin{equation}
    {L_{\rm Ni}(t')\over 10^{42}\rm\,erg/s} \approx \lrb{6.45\mr{e}^{-t'/t_{\rm Ni}} + 1.45 \mr{e}^{-t'/t_{\rm Co}}} {M_{\rm Ni}\over 0.1M_\odot},
\end{equation}
where $M_{\rm Ni}$ is the total mass of $^{56}\mr{Ni}$ and the decay timescales are $t_{\rm Ni} = 8.8\mr{\,d}$ and $t_{\rm Co} = 111.3\rm\,d$. The optical depth $\tau_{\rm MeV}$ accounts for the leakage effect for $\rm \sim\!MeV$ $\gamma$-rays produced in the decays, with
\begin{equation}
    \tau_{\rm MeV} \approx (t'/t_{\rm MeV})^{-2},
\end{equation}
where the leakage time $t_{\rm MeV}$ is obtained from eq. (\ref{eq:gamma_leakage_time}) for an opacity of $\kappa_{\rm MeV}\simeq 0.06\rm\,cm^2\,g^{-1}$ for MeV photons. The second term in the total heating rate, $L_{\rm sh}$, comes from the emission from the shock-heated regions which is assumed to closely track the accretion power $$L_{\rm sh}(t') = \eps_{\rm rad} L(t')$$ with a radiative efficiency of $\eps_{\rm rad}\leq 1$.

To construct a simple model, we consider the \textit{orbit-averaged} accretion power
\begin{equation}
    L(t') = \eps \dot{M}_{\rm P} c^2, \ \ \dot{M}_{\rm P}(t') = \Delta M_{\rm cap}(t')/P_{\rm orb},
\end{equation}
where $\eps$ is the accretion efficiency, $P_{\rm orb}=10\rm\,d$ is the orbital period, $\Delta M_{\rm cap}(t')$  is the per-orbit mass capture at pericenter passage time $t'$, and $\dot{M}_{\rm P}$ is the orbit-averaged mass-capture rate. This form of the accretion power is a very crude approximation (especially at early times), as it does not take into account the discreteness of the accretion episodes that lead to periodic modulation of the lightcurve. We take the radiative efficiency of the disk outflow to be in the following form
\begin{equation}
    \eps_{\rm rad} = {\dot{M}_{\rm P}/\dot{M}_{\rm P,cr} \over 1 + \dot{M}_{\rm P}/\dot{M}_{\rm P,cr}},
\end{equation}
where $\dot{M}_{\rm P,cr} \simeq 10^{-3}\,\msunyr\approx 3\times10^{-5}M_\odot/10 \rm\,d$ is a critical orbit-averaged accretion rate above which the radiative efficiency of internal shocks approaches unity (below $\dot{M}_{\rm P,cr}$, the inverse-Compton cooling becomes less efficient, see \S \ref{sec:shock_emission}).

\section{Application to SN2022jli}\label{sec:application_SN2022jli}

The peculiar transient SN2022jli was initially spectroscopically classified as a Type Ic supernova from the nearby host galaxy NGC 157 at a distance\footnote{The distance to SN2022jli may be as low as 12 Mpc \citep{Chen:2024aa}, which will make all luminosities lower than the values quoted in this paper by a factor of 3. } of $\sim\!22\rm\, Mpc$. The late-time properties of this source are summarized as follows \citep{Moore:2023aa, Chen:2024aa, cartier24_SN2022jli}:

\begin{deluxetable*}{ccccc}
\centering
\tablewidth{0pt}
\tablecaption{Comparison between possible theories for the extra power source in SN2022jli. \label{tab:theories}
}
\tablehead{
  \colhead{observations$\backslash$theories} & \colhead{CSM interaction} &
 \colhead{fallback accretion} &  \colhead{magnetar} & \colhead{companion accretion}}
\startdata
\phn
periodicity  & X  & X  & X & \checkmark \\
embedded power source & nearby CSM? & \checkmark & \checkmark & \checkmark\\
delayed onset ($t\sim 50\rm\,d$) & \checkmark & X & changing $\mu_{\rm B}$? & \checkmark\\
rapid shutoff ($t\sim 270\rm\,d$) & \checkmark & fallback rate cutoff? & X & \checkmark\\
GeV emission & \checkmark & \checkmark & \checkmark & \checkmark\\
H$\alpha$ line & H-rich CSM? & X & evaporating companion? & \checkmark
\enddata
\end{deluxetable*}

\begin{itemize}
    \item[(i)] During the decline phase of normal Type Ic evolution, the optical/near-infrared emission unexpectedly rebrightened and reached a second peak with maximum luminosity of $L\sim 10^{42.5}\rm\, erg\,s^{-1}$ around 50 days since discovery (or roughly 60 days after the inferred explosion time). After a long, gradual decline, the optical and near-infrared luminosities abruptly dropped by more than one order of magnitude around 270 days since discovery.
    This lightcurve behavior requires an additional power source (other than radioactive decay) that turned on with a delay of 1 to 2 months after the explosion and then switched off about 200 days later. 
    \item[(ii)] When the additional power source is on, both the bolometric and monochromatic ($g,r,i,c,o,z$ bands) lightcurves show coherent modulations at a period of $P\simeq 12.4\rm\, d$. The phase-folded lightcurves show a fast-rise and gradual-decline profile, which indicates rapid onset of the energy injection and relatively slow release of the injected energy. The rapid onset timescale of $\Delta t \sim 3\rm\, d$ constrains the physical size of the power source to be $< c \Delta t \sim 10^{16}\rm\, cm$ (for a non-relativistic source). Comparing this with the characteristic radius of the supernova ejecta $R_{\rm ej}\sim v_{\rm ej} t \sim 10^{16}\mr{\,cm}\, (v_{\rm ej}/10^9{\mr{\,cm\,s^{-1}}}) (t/100\rm\, d)$, one infers that the extra power source is embedded inside the ejecta.
    \item[(iii)] During the second lightcurve peak, a narrow hydrogen H$\alpha$ emission line is detected and the temporal variations in the blueshift/redshift (with full-width half-maximum $\Delta v_{\rm FWHM}\sim \rm 1000\, km\,s^{-1}$) of the H$\alpha$ line center follow the same period as in the lightcurve variability. The H$\alpha$ line luminosity stayed roughly 0.4\% of the optical/near-infrared psudo-bolometric luminosity and it had temporal variations consistent with the periodic modulation of the continuum emission. This shows that the narrow H$\alpha$ line is causally connected to the extra power source. For instance, the line emitting gas is likely photoionized by the radiation from the extra power source\footnote{This is supported by the disappearance of the permitted narrow OI emission lines after the shutoff of the extra power source.}. The rapid line profile variability also indicates that the hydrogen gas is embedded inside the (hydrogen-poor) ejecta.
    \item[(iv)] Temporally and spatially coincident high-energy emission in the 1--3 GeV band is detected in a single 2-month time bin between $180\rm\,d$ and $240\rm\,d$ post discovery. The averaged $\gamma$-ray luminosity in this time window is $L_{1\mbox{--}3\rm\, GeV}\simeq 10^{41.5}\rm\, erg\,s^{-1}$, which is a factor of $\sim2$ less than the contemporary optical/near-infrared luminosity. This shows that a significant fraction of the energy in the power source is channeled into non-thermal particles. Moreover, the 12.4-day phase-folded photon arrival times are significantly clustered near the fast-rise segment of the time window (with $\Delta t\sim 3\rm\, d$) and the GeV emission may to be periodically modulated at the statistical significance of 2--3$\sigma$ level (limited by the small number of photons).
\end{itemize}

A number of theories have been considered by the original authors, including (i) interaction between the supernova ejecta with circum-stellar medium (CSM), (ii) accretion of marginally bound fallback material by the compact object, (iii) magnetic dipole spindown of the newborn magnetar (strongly magnetized neutron star), and (iv) accretion from a shock-inflated companion star. In Table \ref{tab:theories}, we list the pros and cons of these models. Here, ``\checkmark'' means that the model prediction is consistent with the observed property, ``X'' means inconsistency, and ``?'' (with possible reasons) means the model can potentially work under special conditions.

Recently, \citet{King:2024aa} proposed a model that is similar to Roche-lobe overflow in ultra-luminous X-ray sources (but with extreme beaming), and they also considered the orbital evolution of the system. Here, for simplicity, we put their model in the (iv) category in Table \ref{tab:theories}, but they did not consider the expansion of the companion star's outer envelope. Another model proposed by \citet{Zhu:2024ab} is based on the gradual evaporation of the companion star by the magnetar's spindown-powered wind and, in their model, a massive hydrogen-rich evaporation flow caused the delayed release of the spindown energy due to a longer photon diffusion timescale. For simplicity, we put their model in the (iii) category in Table \ref{tab:theories} because accretion is not considered. Another possibility for the delayed emission is that the magnetic dipole moment $\mu_{\rm B}$ may gradually increase on a timescale of one to two months \citep{Chugai:2022aa}. However, it is difficult for a model based on the spindown power to reproduce the periodicity in the lightcurve of SN2022jli. The same difficulty applies to the fallback accretion scenario.

We note that CSM interaction can produce shocks that are embedded within the supernova ejecta, as long as part of an equatorial concentrated CSM is located sufficiently close to the exploding star \citep[see e.g., Fig. 2 of][]{smith17_Type_IIn_Ibn}. To reproduce the delayed onset of the energy injection, the inner edge of the CSM is located at $v_{\rm ej} t_{\rm begin}\sim 3\times10^{15}\mr{\,cm}\, (v_{\rm ej}/10^9\mr{\,cm\,s^{-1}}) (t_{\rm begin}/30\rm\,d)$. For the interaction to last for a duration of $t_{\rm end}\sim 270\rm\, d$ (before the rapid shutoff), a fraction of the CSM gas needs to be located at a distance of $v_{\rm ej} t_{\rm end}\sim 2\times10^{16}\mr{\,cm}\, (v_{\rm ej}/10^9\mr{\,cm\,s^{-1}}) (t_{\rm end}/270\rm\,d)$. A potential difficulty of the extended hydrogen-rich CSM is that photoionization by the supernova emission may produce narrow H$\alpha$ lines that are not detected in the early time spectra.
A more serious difficulty of the CSM-interaction picture lies on the nearly constant period in the observed lightcurve. Even if the CSM has a spatially periodic structure (i.e., the density is a periodic function of radius), a time-dependent outflow speed can easily destroy the periodicity in the shock power.

In the following, we show that the model based on accretion from the shock-inflated companion star can reproduce the main properties of SN2022jli (\S \ref{sec:orbit_constraints}--\ref{sec:Halpha_line}).

\subsection{Periodicity and orbital constraints}\label{sec:orbit_constraints}


In this model, the periodic modulation in the lightcurve is due to orbital motion. Here, we consider the orbit of the system before and after the supernova. The post-supernova semimajor axis is given by
\begin{equation}\label{eq:postSN_sma}
\begin{split}
    a &= \lrsb{G(M_{\rm ns} + M_*)\lrb{P_{\rm orb}\over 2\pi}^2}^{1/3} \\
    &\approx 36 R_\odot \lrb{M_{\rm ns} + M_*\over 4M_\odot}^{1/3} \lrb{P_{\rm orb}\over 12.4\mr{\,d}}^{2/3},
\end{split}
\end{equation}
where we have ignored the small mass loss from the companion star.

The total binary mass before the supernova is
\begin{equation}
    M_{\rm tot} = M_{\rm ns} + M_* + M_{\rm ej},
\end{equation}
where $M_{\rm ej}$ is the ejecta mass. We ignore the small difference between the baryonic and gravitational masses of the neutron star. For a pre-supernova circular orbit with semi-major axis $a_0$, the relative (orbital) velocity between the two stars is
\begin{equation}
    v_{\rm orb,0} = \sqrt{G(M_{\rm ns} + M_* + M_{\rm ej})/a_0}.
\end{equation}
Let us adopt a coordinate system where the pre-supernova orbit is in the xy plane. We ignore the small momentum deposition on the companion star by the supernova ejecta \citep[see][for a thorough discussion]{Hirai:2018aa}. In the instantaneous comoving frame of the companion star, the supernova progenitor star is located at $a_0 \hat{x}$ and is moving in the $\hat{y}$ direction with velocity $\vec{v}_{\rm orb,0} = v_{\rm orb,0} \hat{y}$. After the supernova, the neutron star receives a natal kick which is denoted as $\vec{v}_{\rm k}$, so the total specific energy (per reduced mass) of the post-supernova orbit is given by
\begin{equation}
    E_{\rm orb} = (\vec{v}_{\rm orb,0} + \vec{v}_{\rm k})^2/2 - G(M_{\rm ns} + M_*)/a_0.
\end{equation}
If $E_{\rm orb} < 0$, then the binary stays bound with a new semimajor axis $a$ that is given by
\begin{equation}
    E_{\rm orb} = -{G(M_{\rm ns} + M_*)\over 2a},
\end{equation}
and the post-supernova orbital period $P$ is related to $a$ by eq. (\ref{eq:postSN_sma}).
Regardless of the boundness of the orbit, the pericenter separation is given by ($a$ may be negative in this expression if $E_{\rm orb} > 0$)
\begin{equation}
    r_{\rm p} = a\lrb{1 - \sqrt{1 - {\ell^2\over (M_{\rm ns} + M_*) a}}},
\end{equation}
where the specific angular momentum (per reduced mass) is given by
\begin{equation}
    \vec{\ell} = a_0\hat{x} \times (v_{\rm orb,0} \hat{y} + \vec{v}_{\rm k}).
\end{equation}
For the unbound cases, there is a 50\% chance that the neutron star actually reaches the orbital pericenter. If the post-supernova total velocity is pointing away from the companion star, the neutron star will not reach the pericenter (in such cases, the closest separation between the NS and the companion star is $a_0$). Generally, we have the following constraint on the post-supernova orbit:
\begin{equation}
    r_{\rm p} \leq a_0.
\end{equation}



\begin{figure}
\centering
\includegraphics[width=0.47\textwidth]{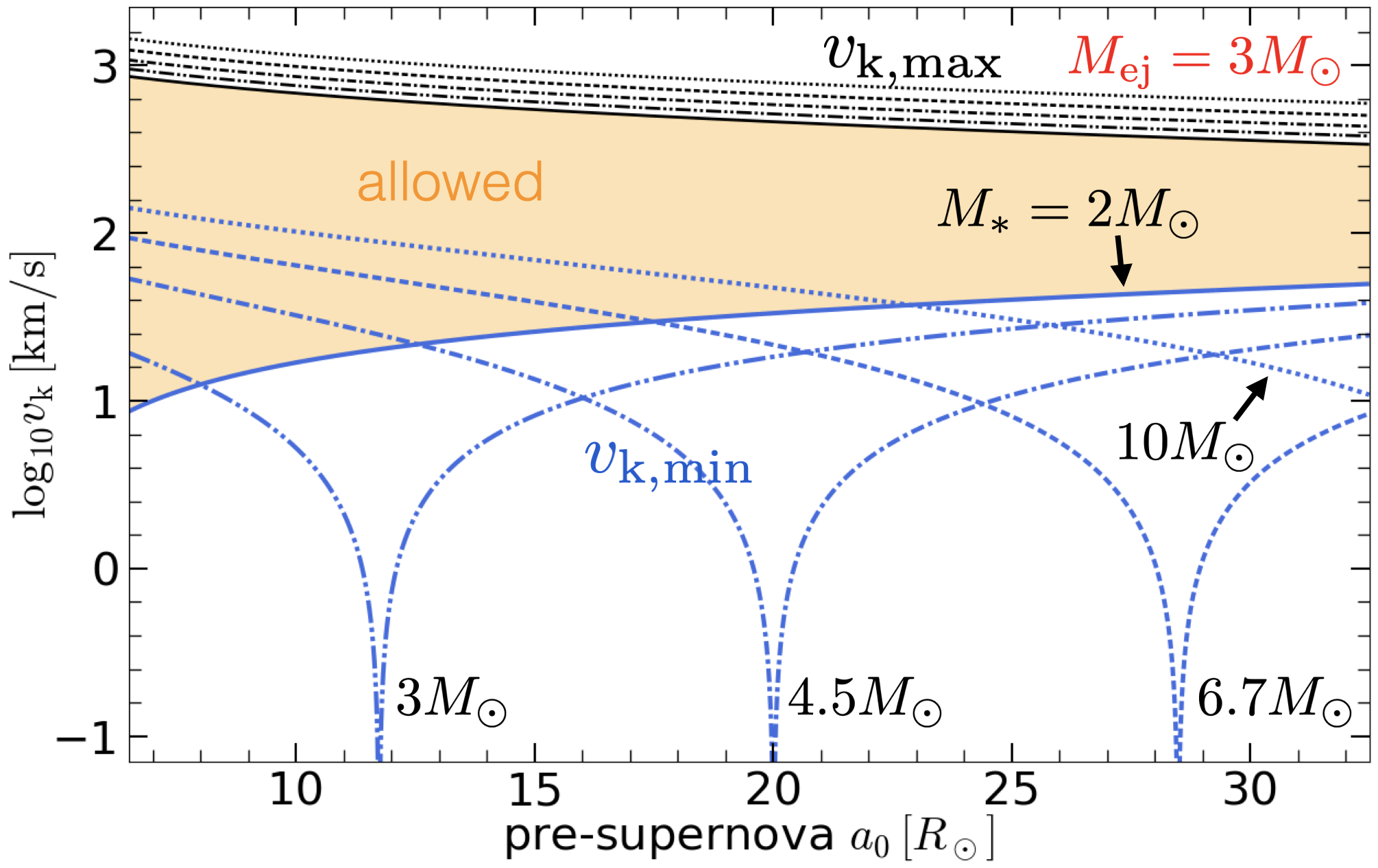}
\caption{Constraints on the magnitude of the natal kick $v_{\rm k,min} < v_{\rm k} < v_{\rm k,max}$. The thin black lines are for the maximum kicks $v_{\rm k,max}$ and the thick blue lines are for the minimum $v_{\rm k,min}$. For each companion star mass $M_*$ and pre-supernova separation $a_0$, the allowed $v_{\rm k}$ lies in between the two limits. We marked the allowed region for the $M_*=3M_\odot$ case in orange and the allowed region is different for other $M_*$. The ejecta mass is fixed to be $M_{\rm ej} = 3M_\odot$ for all cases, but other ejecta masses $1\lesssim M_{\rm ej}\lesssim 10M_\odot$ give qualitatively similar results.
}
\label{fig:vk_constraints}
\end{figure}

The magnitude of the natal kick may be constrained by the post-supernova orbital period $P_{\rm orb}=12.4\rm\, d$. For a given kick magnitude $v_{\rm k} = |\vec{v}_{\rm k}|$, $P$ is the longest (or shortest) when $\vec{v}_{\rm k}$ is aligned (anti-aligned) with $\vec{v}_{\rm orb,0}$. Thus, for a known orbital period, we obtain a lower limit and an upper limit for the kick amplitude
\begin{equation}
\begin{split}
    v_{\rm k,min} &= \abslt{\sqrt{G(M_{\rm ns} + M_*) (2/a_0 - 1/a)} - v_{\rm orb,0}}, \\
    v_{\rm k,max} &= \sqrt{G(M_{\rm ns} + M_*) (2/a_0 - 1/a)} + v_{\rm orb,0}.
\end{split}
\end{equation}
These limits are shown in Fig. \ref{fig:vk_constraints}. Unfortunately, we only find weak constraints on the kick amplitude: the region with $100 \lesssim v_{\rm k}\lesssim 500\rm\, km\,s^{-1}$ is generally allowed, for any companion star masses in between 1 and 10 $M_\odot$. Although $v_{\rm k,min}$ may be as low as zero for a particular combination of $M_{\rm ej}, M_*$ and $a_0$, we believe this to be a fine-tuned condition. Instead, a finite, minimum kick is statistically required, but the constraint is rather weak ($v_{\rm k,min}\gtrsim 10\rm\, km\,s^{-1}$). We conclude that a wide range of kick velocities between 10 and 500 $\rm km\,s^{-1}$ may potentially reproduce $P=12.4\rm\,d$. This simply means that it is rather natural to produce a post-supernova orbital period of the order 10 days.

\subsection{Delayed onset and rapid shutoff}\label{sec:onset_shutoff}

In our model, the delayed onset of the second peak of the optical lighturve at $t\sim 50\rm\,d$ is due to slow expansion of the companion star after the shock impact --- this occurs on a timescale of $t_{\rm max}$ (eq. \ref{eq:tmax_companion_star}). The rapid shutoff of the lightcurve after $t\sim 270\rm\, d$ is caused by the Kelvin-Helmholtz contraction of the stellar envelope. In this subsection, we use the timing of the delayed onset and rapid shutoff to constrain the pre-supernova orbital separation $a_0$, whereas the constraint on the companion star's mass is rather weak.

Since the inflation of the shock-heated companion star takes place on a timescale of $t_{\rm max}$, which mainly depends on the pre-supernova orbital separation $a_0$ (eq. \ref{eq:tmax_companion_star}), we use $50\lesssim t_{\rm max}\lesssim 270\rm\, d$ to constrain $a_0$. Based on Fig. \ref{fig:Rmax_tmax} (blue shaded region in the upper panel), we obtain the following conservative constraints on $a_0$:
\begin{equation}
    10\lesssim a_0\lesssim 30R_\odot, \mbox{ (from SN2022jli timing)}
\end{equation}
which is only very weakly dependent of companion mass. This is consistent with the sufficient condition for accretion (eq. \ref{eq:condition_for_accretion}), as long as the pre-supernova orbital separation is $a_0\lesssim 20R_\odot$. 

However, in our simplified lightcurve model (Fig. \ref{fig:lightcurve}), the onset and shutoff of the accretion-powered second peak are not as sharp as observed.

We speculate that this might be due to the effects of accretion feedback on the structure of the star's outer envelope. At early time when $R_*$ just expands past $\rp$, the outermost layers of the envelope may be rapidly removed by the mechanical energy output from accretion. This is possible for sufficiently strong ram pressure of the disk outflow $P_{\rm w}\gtrsim 10^8\rm\, erg/cm^3$ (eq. \ref{eq:disk_wind_pressure}), which would overwhelm the ram pressure of the envelope gas in the neutron star's frame (eq. \ref{eq:ram_pres_accretion_flow}). It is possible that the accretion onto the neutron star is only strong when the envelope's density at $r = r_{\rm p}$ exceeds $10^{-7}\rm\, g\,cm^{-3}$. Modeling the interactions between the disk outflow and the stellar envelope is beyond the scope of this paper.

Another possibility is that when the accretion rate drops below a threshold given by $P=P_{\rm cr}$ (eq. \ref{eq:Omega_cr}), the propeller wind suddenly disappears and the total power of the system drops by a factor of $\sim$10 (eq. \ref{eq:Lm_over_Lw_at_Pcr}. This could potentially give rise to more abrupt onset and cutoff in the accretion power and hence the lightcurve in the second peak. Due to theoretical uncertainties of the propeller wind, we do not discuss this possibility in detail.

\subsection{GeV emission}\label{sec:GeV_emission}

In our model, the GeV emission is produced by $pp$ collisions between the cosmic ray protons accelerated by internal shocks ($r_{\rm sh}\sim10^{14}\rm\, cm$) and the non-relativistic thermal protons. To explain the observed average luminosity of $L_{\rm 1\mbox{--}3\,GeV}\sim 3\times10^{41}\rm\, erg\,s^{-1}$ in the 180-240 d time bin, our model likely requires a combination of a relatively high cosmic ray energy efficiency ($\eps_p > 0.1$) and a large per-orbit accretion energy release $\eps \Delta M_{\rm cap} c^2 > 2\times10^{48}\rm\, erg$. Preliminary hydrodynamic numerical simulations (Cary et al., in prep) indeed suggest that $\Delta M_{\rm cap}$ is higher than our estimate of $10^{-4}M_\odot$ by a factor of a few to 10, so it is plausible to produce the observed GeV luminosity.

More importantly, the delayed onset of the GeV emission means that GeV photons are absorbed by Bethe-Heitler pair production process at $t\lesssim 150\rm\, d$, where the upper limit in the escaping time for GeV photons is conservative considering the poor time resolution of the Fermi-LAT lightcurve. Since the opacity for GeV photons is $\kappa_{\rm GeV}\simeq 0.03\rm\, cm^2\,g^{-1}$, we infer from eq. (\ref{eq:gamma_leakage_time}) a relatively large ejecta mass of $M_{\rm ej}\simeq 3M_\odot$.
A potential, independent measurement of the ejecta mass may be obtained from the nebular phase ($t\gtrsim 300\rm\, d$) of the lightcurve evolution.

\subsection{H$\alpha$ line}\label{sec:Halpha_line}

A natural prediction of our model is the production of hydrogen recombination lines such as H$\alpha$ due to photo-ionization of the slowest (and densest) parts of the disk wind that has not been shock-heated.

Depending on whether the Str\"omgren sphere encloses the entire wind, the system may be in one of the two regimes: ionization-limited (where a neutral layer exists) or density-limited (the wind is fully ionized). The maximum ionization rate is given by $\dot{N}_{\rm ion} \sim \xi L_{\rm sh}/\mc{E}_{\rm ion}\sim 10^{53}{\rm\, s^{-1}}\, L_{\rm sh,43}$, where we have taken a fraction $\xi\sim 0.5$ of the luminosity $L_{\rm sh}$ from internal shocks to be used in ionization and each ionization costs an average amount of energy $\mc{E}_{\rm ion}\sim 30\rm\, eV$ (the ionization is physically due to Compton-heated electrons colliding with hydrogen atoms). The maximum mass of the ionized gas is
\begin{equation}
    M_{\rm ion,max}\sim {\dot{N}_{\rm ion} m_p\over n \alpha_{\rm B}}\sim 10^{-2}M_\odot \dot{N}_{\rm ion,52} n_{11}^{-1} T_5^{0.7},
\end{equation}
where $n$ is the gas number density (eq. \ref{eq:proton_density_shocked_region}), $\alpha_{\rm B}\simeq 2.6\times10^{-13}(T/10^4\rm\, K)^{-0.7}$ is the Case-B recombination coefficient, and we have taken $T=10^5T_5\rm\,K$ as a fiducial value for the temperature of the ionized gas. We find $M_{\rm ion,max} > \Delta M_{\rm cap}$ to be the typical case, so the system is in the density-limited regime, which leads to a relatively small H$\alpha$ yield $L_{\rm H\alpha}/L_{\rm sh}< 1\%$ than the ionization-limited regime. From the H$\alpha$ recombination branching fraction $\alpha_{\rm B}^{\rm H\alpha}/\alpha_{\rm B}\approx 0.3$, we can estimate the H$\alpha$ luminosity by assuming that the entire slow wind of mass $\Delta M_{\rm cap}$ is fully ionized,
\begin{equation}
\begin{split}
    L_{\rm H\alpha} &\sim n \alpha_{\rm B} {\Delta M_{\rm cap}\over m_p} {\alpha_{\rm B}^{\rm H\alpha}\over \alpha_{\rm B}} \mc{E}_{\rm H\alpha}\\
    &\sim 10^{39}\mr{erg\,s^{-1}} n_{11} T_5^{-0.7} {\Delta M_{\rm cap}\over 10^{-4}M_\odot},
\end{split}
\end{equation}
where $\mc{E}_{\rm H\alpha}=1.9\rm\, eV$ is the H$\alpha$ photon energy.
Since the gas density scales as $n\propto \Delta M_{\rm cap}/r_{\rm sh}^{-3}$, we find $L_{\rm H\alpha}\propto \Delta M_{\rm cap}^2/r_{\rm sh}^{-3}$. The observed H$\alpha$ luminosity reaches up to $5\times 10^{39}\rm\, erg\,s^{-1}$ (and possibly even higher near the 2nd peak of the optical lightcurve), which can be produced if the Bondi-captured mass per orbit is $\Delta M_{\rm cap}\gtrsim 3\times10^{-4}M_\odot$.

Another piece of information is the H$\alpha$ line profile, which is observed to undergo rapid variations of the Doppler shift on the orbital timescale. The peak of the line has velocities of the order $10^3\rm\, km\,s^{-1}$, but the width may extend to larger velocities up to $5000\rm\, km\,s^{-1}$. These velocities are consistent with our model where the H-rich disk outflow has a wide range of velocities and most of the mass is contained in the slowest outflow components near $10^3\rm\, km\,s^{-1}$ that dominates the H$\alpha$ luminosity.

\section{Summary}\label{sec:summary}

In this work, we study the aftermath of the interaction between the ejecta of a stripped-envelope supernova and a close-by main-sequence companion star, motivated by the recent observations of the peculiar Type Ic supernova SN2022jli \citep{Moore:2023aa, Chen:2024aa, cartier24_SN2022jli}.

We first construct a 1D model for the shock-heated stellar envelope by entropy injection according to the Rankine-Hugoniot jump condition. Then, we use MESA to follow the subsequent evolution of the ejecta-impacted star. It is found that the stellar envelope gradually expands to a maximum radius $R_{\rm max}$ up to $\sim\! 10^2R_\odot$ (eq. \ref{eq:Rmax_companion_star}) that is mainly dependent on the pre-supernova separation $a_0$. If the neutron star remains in a bound orbit, the pericenter radius $\rp$ is constrained to be $\rp \leq a_0$. Thus, a sufficient condition for neutron star accretion from the stellar envelope is given by $R_{\rm max}(a_0)>a_0$, which is shown to correspond to $a_0\lesssim 20\rm\, R_\odot$ (nearly independent of the companion star's mass) or pre-supernova orbital period $P_0\lesssim 3\mr{\,d} \,(M_{\rm tot}/10M_\odot)^{-1/2}$ for a total mass of $M_{\rm tot}$. Neutron star accretion may also occur for wider pre-supernova orbits for favorable orientations of the natal kick.

The time it takes for the ejecta-impacted companion star to expand to the maximum radius also mainly depends on $a_0$ as $t_{\rm max}\sim 290\mr{\,d}\, (a_0/10R_\odot)^{-12/5}$ (eq. \ref{eq:tmax_companion_star}). The slow expansion of the stellar envelope leads to a delayed onset of the neutron star accretion, which may cause the supernova optical lightcurve to be double-peaked, as the onset of the accretion power is usually delayed by more than the peak time of Type Ib/c supernovae ($\sim\!10\rm\, d$). At late time $t > t_{\rm max}$, the outer envelope undergoes Kelvin-Helmholtz contraction, and this causes the accretion power to disappear, so the supernova emission returns to the normal nebular phase.

We then study the accretion power from the gravitationally captured gas, taking into account the interactions between the accretion disk and the neutron star magnetosphere. We find that, for $a_0\lesssim 20\rm\, R_\odot$, the typical mass capture per orbit is $\Delta M_{\rm cap}\gtrsim 10^{-4}\rm\, M_\odot$ near the maximum stellar radius ($R_*\sim R_{\rm max}$). The disk has a high accretion rate of the order $\dot{M}_{\rm d}\sim 0.1M_\odot/\rm yr$, and a powerful disk wind may be produced from near the Alfv\'en radius where the disk is truncated by the neutron star magnetospheric pressure. We find that an accretion efficiency of the order $\eps = \Delta E/(\Delta M_{\rm cap}c^2) \sim 10^{-2}$ is possible when the system is in the ``propeller'' state for a typical Crab-like neutron star. Thus, each accretion episode produces an energy of the order $\Delta E \sim 10^{48}{\rm\, erg}\, \eps_{-2} (\Delta M_{\rm cap}/10^{-4}M_\odot)$, mostly in the kinetic form. 

The disk wind has a wide range of velocities from $v_{\rm min}\sim 10^3\rm\, km/s$ to a fraction of speed of light, as the slowest parts (carrying most of the mass $\Delta M_{\rm cap}$) are launched from the outer disk near $\rd\sim 0.3 R_\odot$ (eq. \ref{eq:disk_radius}) and the fastest parts (carrying most of the energy $\Delta E$) are from the inner disk near the Alfv\'en radius. A consequence of the velocity stratification is that internal shocks are produced near the characteristic radius $r_{\rm sh}\sim v_{\rm min} P\sim 10^{14}\rm\, cm$ for an orbital period of $P\sim 10\rm\,d$. We find the internal shocks to be radiative, thanks to efficient inverse-Compton cooling. Thus, a large fraction of $\Delta E$ will be converted into radiation energy (mostly carried by MeV photons), which will then be reprocessed by the supernova ejecta into the optical band. The unshocked slowest parts of the disk wind are photo-ionized, which produces narrow ($\sim\!10^3\rm\, km/s$) Balmer lines by recombination that would otherwise be highly unusual for Type Ib/c supernovae in terms of the hydrogen composition and linewidth.

We also study the non-thermal emission from the internal shocks. It is found that cosmic ray protons rapidly cool due to hardronic $pp$ collisions, and hence a fraction $\eps_p\sim 0.1$ of the outflow energy $\Delta E$ is converted into $\gamma$-rays and neutrinos --- these high-energy messengers have energies ranging from 100 MeV up to 10 PeV, but most of their energies are carried by $\sim\!\rm GeV$ ones that are produced by mildly relativistic cosmic ray protons, with a typical time-averaged luminosity of $L_{\rm GeV}\sim 10^{41}\rm\, erg\,s^{-1}$ (eq. \ref{eq:GeV_luminosity}) that is comparable between GeV photons and neutrinos. The neutrinos escape without any obscuration, but GeV photons are absorbed by the supernova ejecta in the first $\sim\!100\rm\, d$ (eq. \ref{eq:gamma_leakage_time}) due to Bethe-Heitler pair production. This causes the emerging GeV emission to be delayed compared to the 2nd peak of the optical lightcurve. 

Our model provides a self-consistent explanation of the puzzling properties of SN2022jli, including the delayed onset and rapid shutoff of the second peak, periodic modulations, delayed GeV emission, and narrow Balmer lines. The main requirement is that the companion star was at a very close pre-supernova separation of $10\lesssim a_0\lesssim 20R_\odot$ (or orbital period $1\lesssim P_0\lesssim 3\rm\, d$ for a total mass near $10M_\odot$), which comes from the requirement of $50\lesssim t_{\rm max}\lesssim 270\rm\, d$. Such a tight binary system is most likely formed by common-envelope evolution \citep[as found in population synthesis models, e.g.,][]{ko25_binary_pop_syn}.
Unfortunately, our model does not provide a strong constraint on the mass of the companion star. It may be possible to constrain the companion mass by directly observing the companion star after the supernova has faded away. A comparison between Figs. \ref{fig:evolve_shocked_star_one_2Msun_HR} and \ref{fig:evolve_shocked_star_one_10Msun_HR} suggests that a higher mass companion star would appear like a blue supergiant for a longer duration. For instance, a star with $L_*\sim 10^5L_\odot$ and $T_{\rm eff}\simeq 10^4\rm\, K$ would have an AB magnitude of 24-25 mag in the optical bands at a distance of 22 Mpc. However, detecting such a star would require difference imaging with high angular resolution.

Our model motivates detailed hydrodynamic simulations of the interactions between a newly born neutron star and a shock-inflated companion star, and the internal shocks between adjacent episodes of disk outflows.

\section*{Acknowledgment}

We thank Ping Chen, J.J. Eldridge, and Raffaella Margutti for useful discussions. The research of W.L. and S.C. is supported by the Hellman Fellows Fund. This research benefited from interactions at workshops funded by the Gordon and Betty Moore Foundation through grant GBMF5076 and through interactions at the Kavli Institute for Theoretical Physics, supported by NSF PHY-2309135. D.T. is supported by the Sherman Fairchild Postdoctoral Fellowship at Caltech.

\bibliographystyle{mnras}
\bibliography{ref.bib}

\appendix

\begin{figure*}
\centering
\includegraphics[width=0.8\textwidth]{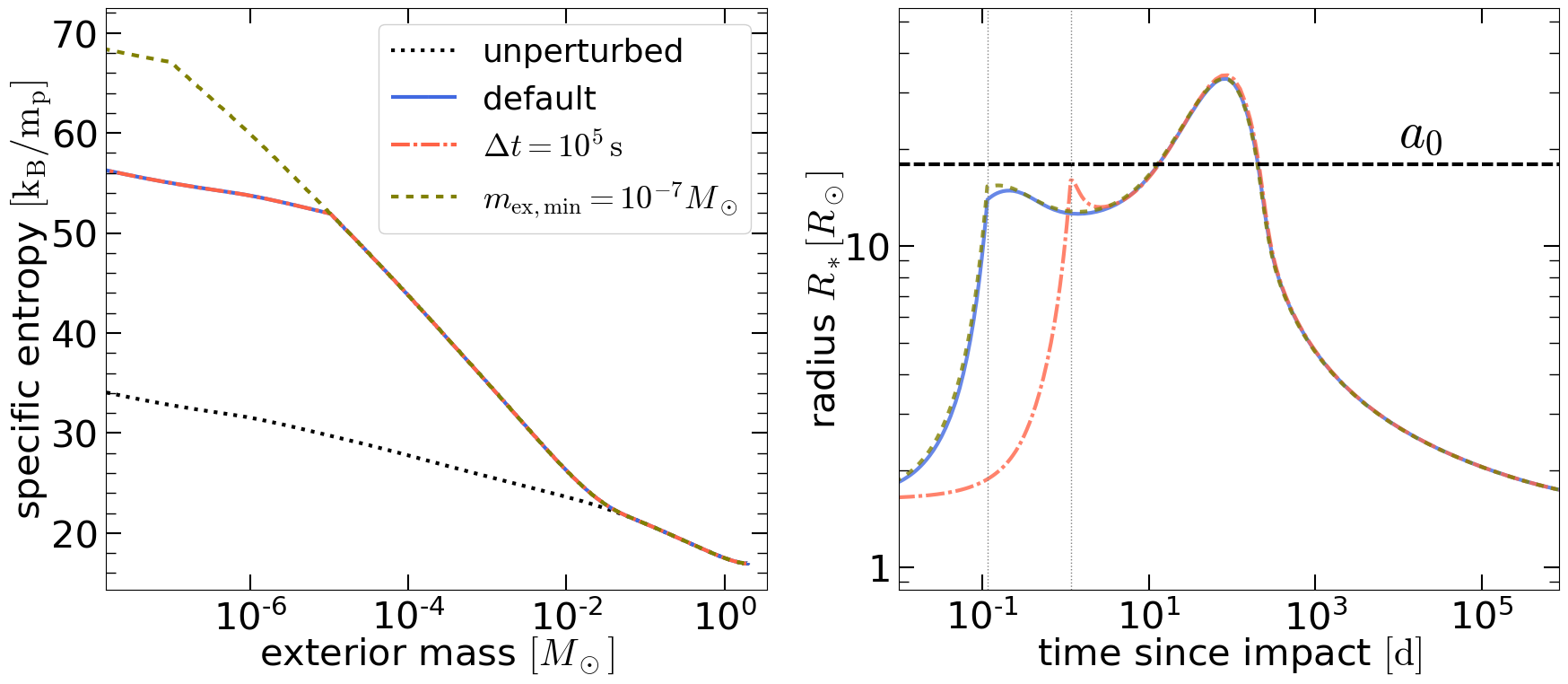}
\caption{The left panel shows the entropy profiles before (black dotted) and after (colored lines) shock heating. The right panel shows the radius evolution of the star in different cases. The default case (blue solid line) of has heating time $\Delta t=10^4\rm\, s$ and minimum exterior mass $m_{\rm ex,min}=10^{-5}M_\odot$ below which the entropy jump is taken to be flat. Two other cases are considered: one with $\Delta t=10^5\rm\, s$ (red dash-dotted line) and the other one (green dashed line) with $m_{\rm ex,min}=10^{-7}M_\odot$. For all cases, we take $E_{\rm ej}=10^{51}\rm\, erg$, $M_*=2M_\odot$, and pre-supernova separation $a_0=18R_\odot$.
}
\label{fig:test_ds_flattening}
\end{figure*}

In this appendix, we show Fig. \ref{fig:test_ds_flattening} which demonstrates that our results are insensitive to the choices of envelope heating time $\Delta t$ (eq. \ref{eq:heating_rate}) and minimum exterior mass $m_{\rm ex,min}$ (eq. \ref{eq:mex_min}).

\end{document}